\documentclass[final,1p,times]{elsarticle}

\usepackage{bm}
\usepackage{color}
\usepackage{array}
\usepackage{graphicx}
\usepackage{multirow}
\usepackage{booktabs}
\usepackage{fancyvrb}
\usepackage{upquote}

\usepackage{geometry}

\newcolumntype{P}[1]{>{\centering\arraybackslash}p{#1}}
\newcolumntype{M}[1]{>{\centering\arraybackslash}m{#1}}

\newcommand{\be}{\begin{equation}}
\newcommand{\ee}{\end{equation}}
\newcommand{\een}{\end{subequations}}
\newcommand{\ben}{\begin{subequations}}
\newcommand{\beq}{\begin{eqalignno}}
\newcommand{\eeq}{\end{eqalignno}}

\newcommand{\lsim}{\mathrel{\mathop{\kern 0pt \rlap
      {\raise.2ex\hbox{$<$}}}\lower.9ex\hbox{\kern-.190em $ \sim$}}}
\newcommand{\gsim}{\mathrel{\mathop{\kern 0pt
      \rlap{\raise.2ex\hbox{$>$}}}\lower.9ex\hbox{\kern-.190em $\sim$}}}

\newcommand{\op}{{\cal O}}

\newcommand{\C}{{\cal C}}
\newcommand{\Q}{{\cal Q}}

\newcommand{\todo}[1]{{\color{red} \ifmmode\else[todo]\fi #1}}

\usepackage{amssymb}

\journal{Astroparticle Physics}

\begin{document}

\begin{frontmatter}



\title{On the sensitivity of present direct
  detection experiments to WIMP--quark and WIMP--gluon effective
  interactions: a systematic assessment and new model--independent approaches.}


\author{Sunghyun Kang}
\ead{francis735@naver.com}
\author{Stefano Scopel}
\ead{scopel@sogang.ac.kr}
\author{Gaurav Tomar}
\ead{tomar@sogang.ac.kr}
\author{Jong--Hyun Yoon}
\ead{jyoon@sogang.ac.kr}
\address{Department of Physics, Sogang University, 
Seoul, Korea, 121-742}

\begin{abstract}
Assuming for Weakly Interacting Massive Particles (WIMPs) a Maxwellian
velocity distribution in the Galaxy we provide an assessment of the
sensitivity of existing Dark Matter (DM) direct detection (DD)
experiments to operators up to dimension 7 of the relativistic
effective field theory describing dark matter interactions with quarks
and gluons . In particular we focus on a systematic approach,
including an extensive set of experiments and large number of
couplings, both exceeding for completeness similar analyses in the
literature. The relativistic effective theory requires to fix one
coupling for each quark flavor, so in principle for each different
combination the bounds should be recalculated starting from direct
detection experimental data. To address this problem we propose an
approximate model--independent procedure that allows to directly
calculate the bounds for any combination of couplings in terms of
model--independent limits on the Wilson coefficients of the
non--relativistic theory expressed in terms of the WIMP mass and of
the neutron--to--proton coupling ratio $c^n/c^p$. We test the result
of the approximate procedure against that of a full calculation, and
discuss its possible pitfalls and limitations.  We also provide a
simple interpolating interface in Python that allows to apply our
method quantitatively.
\end{abstract}


\end{frontmatter}



\section{Introduction}
\label{sec:introduction}

One of the most popular scenarios for the Dark Matter (DM) which is
believed to contribute to up to 27\% of the total mass density of the
Universe~\cite{planck} and to more than 90\% of the halo of our Galaxy
is provided by Weakly Interacting Massive Particles (WIMPs) with a
mass in the GeV-TeV range and weak--type interactions with ordinary
matter.  Such small but non--vanishing interactions can drive WIMP
scattering events off nuclear targets, and the measurement of the ensuing
nuclear recoils in low--background detectors (direct detection, DD)
represents the most straightforward way to detect them.  Indeed, a
large worldwide effort is currently under way to observe WIMP-nuclear
scatterings, but, with the exception of the DAMA collaboration
\cite{dama_1998, dama_2008,dama_2010,dama_2018} that has been
observing for a long time an excess compatible to the annual
modulation of a DM signal, many other experiments using different
nuclear targets and various background--subtraction
techniques~\cite{xenon_2018, panda_2017, kims_2014, cdmslite_2017,
  super_cdms_2017, coupp, picasso, pico60_2015, pico60_2019, cresst_II,
  cdex, damic, ds50} have failed to observe any WIMP signal so far.

The calculation of DD expected rates is affected by large
uncertainties, of both astrophysical and particle--physics nature.
For instance, most of the explicit ultraviolet completions of the
Standard Model that stabilize the Higgs vacuum contain WIMP exotic
states that are viable DM candidates and for which detailed
predictions for WIMP--nuclear scattering can be worked out, leading in
most cases to either a Spin Independent (SI) cross section
proportional to the square of the target mass number, or to a
Spin--Dependent (SD) cross section proportional to the product of the
WIMP and the nucleon spins. Crucially, this allows to determine how
the WIMP interacts with different targets, and to compare in this way
the sensitivity of different detectors to a given WIMP candidate, with
the goal of choosing the most effective detection strategy. However,
the non--observation of new physics at the Large Hadron Collider (LHC)
has prompted the need to go beyond such top--down approach and to use
either ``effective'' or ``simplified'' models to analyze the
data~\cite{queiroz}, implying a much larger range of possible scaling
laws of the WIMP--nucleon cross section on different targets.
Moreover, the expected WIMP--induced scattering spectrum depends on a
convolution on the velocity distribution $f(\vec{v})$ of the incoming
WIMPs, usually described by a thermalized non--relativistic gas
described by a Maxwellian distribution whose root--mean--square
velocity $v_{rms}\simeq$ 270 km/s is determined from the galactic
rotational velocity by assuming equilibrium between gravitational
attraction and WIMP pressure.  Indeed, such model, usually referred to
as Isothermal Sphere, is confirmed by numerical
simulations~\cite{kelso}, although the detailed merger history of the
Milky Way is not known, allowing for the possibility of the presence
of sizable non--thermal components for which the density, direction
and speed of WIMPs are hard to predict~\cite{green}.

As far as the latter issue is concerned, for definiteness, in the
following we will adopt for the velocity distribution $f(\vec{v})$ of
the incoming WIMPs a standard thermalized non--relativistic gas
described by a Maxwellian distribution.

On the other hand, in the present paper we wish to focus on the former
issue of the scaling law in direct detection, and in particular on how
to compare the sensitivities of different experimental set--ups on
WIMP--quark and WIMP--gluon effective interactions by making use of
model--independent bounds obtained independently at a lower scale on
WIMP--nucleon non--relativistic operators. In particular, since the DD
process is non--relativistic (NR) it has been understood some time
ago~\cite{haxton1,haxton2} that the most general interaction besides
the SI and the SD cross sections can be parameterized with an
effective Hamiltonian that complies with Galilean symmetry, containing
at most 15 terms in the case of a spin--1/2 particle:

\begin{eqnarray}
{\bf\mathcal{H}}({\bf{r}})&=& \sum_{\tau=0,1} \sum_{j=1}^{15} c_j^{\tau} \mathcal{O}_{j}({\bf{r}}) \, t^{\tau} ,
\label{eq:H}
\end{eqnarray}

\noindent where the $\mathcal{O}_{j}$ operators are listed in
\cite{haxton2} and $t^0=1$, $t^1=\tau_3$ denote the $2\times2$
identity and third Pauli matrix in isospin space, respectively, and
the isoscalar and isovector coupling constants $c^0_j$ and $c^{1}_j$,
are related to those to protons and neutrons $c^{p}_j$ and $c^{n}_j$
by $c^{p}_j=c^{0}_j+c^{1}_j$ and $c^{n}_j=c^{0}_j-c^{1}_j$.

Indeed, the NR couplings $c_j^{\tau}$ represent the building blocks of
the low--energy limit of any ultraviolet theory, so that an
understanding of the behaviour of such couplings is crucial for the
interpretation of more general scenarios.  As a consequence, the NR
effective theory (NREFT) of Eq.~(\ref{eq:H}) has been extensively used
in the literature to analyze direct detection
data~\cite{chang_momentum_dependence_2010, dobrescu_nreft, fan_2010,
  hill_solon_nreft, peter_nreft, cirelli_tools_2013,
  effective_wimps_2014, catena_nreft,
  catena_directionality_nreft_2015, Catena_Gondolo_global_fits,
  cerdeno_nreft, Catena_Gondolo_global_limits, nreft_bayesian,
  xenon100_nreft, cresst_nreft}.

In light of this, in Ref.~\cite{sogang_scaling_law_nr} we provided an
assessment of the overall present and future sensitivity of an
extensive list of both present and future WIMP direct detection
experiments assuming systematically dominance of one of the possible
terms of the NR effective Hamiltonian in the calculation of the
WIMP--nucleon cross section. In particular, compared to previous
analyses adopting the same approach, in
Ref.~\cite{sogang_scaling_law_nr} the bounds on the NREFT are
presented in a novel model--independent way: for each of the couplings
of Eq.~(\ref{eq:H}) a contour plot of the most stringent 90\%
C.L. bound on the WIMP--nucleon cross section among a comprehensive
set of 14 existing experiments is provided as a function the WIMP mass
$m_{\chi}$ and of the ratio of the WIMP--neutron and WIMP--proton
couplings $c^n/c^p$ (along with a color code showing the experiment
providing it). This approach allows to make the best constraints on
the WIMP--proton cross section available in a model--independent way
(in~\ref{app:program} a simple code is introduced that allows to
interpolate the $m_{\chi}$--$c^n/c^p$ planes of
Ref.~\cite{sogang_scaling_law_nr} to get the corresponding numerical
values) so that, with the exception of cancellations among different
NR operators, such bounds can be directly used to get constraints for
a given relativistic effective DM scenario when taking its
non--relativistic limit, without the need to go through the
calculation of the experimental bound starting from the data and to
apply the standard machinery used in~\cite{sogang_scaling_law_nr}. The
latter includes more refined treatments beside a simple comparison
between theoretical predictions and upper bounds, such as background
subtraction or the optimal-interval method~\cite{yellin}, and may not
be trivial for model builders, who have only access from experimental
papers to the bounds on the standard isoscalar spin--independent or
WIMP--proton/WIMP--neutron spin--dependent cross sections.

However, in spite of its generality, such approach presents some
drawbacks: in particular, the interference of different NR operators
and especially the sensitivity of such effect to the running of the
couplings from the energy scale of the ultraviolet theory to the
nucleon scale~\cite{deramo_2014,deramo_2016,belyaev_reft} are
difficult to include in a model--independent way, as well as a
possible momentum dependence of the Wilson coefficients of the NR
theory. In particular, the latter can arise in the case of a
long--range interaction such as for electric--dipole or
magnetic--dipole DM~\cite{mohanty_2009,anapole_2014}. Moreover an
additional momentum dependence arises when one needs to include the
light-meson poles in the case the DM couples to the axial quark
current \cite{vogel_review,directdm}.

For the reasons listed above, the use of the limits on the NR
couplings of Ref.~\cite{sogang_scaling_law_nr} to calculate the bounds
for an effective relativistic model defined at a much larger scale
needs to be tested. This is the first main goal of the present paper,
where we wish to use the results of Ref.~\cite{sogang_scaling_law_nr}
to calculate the bounds on a specific example of relativistic
effective theory, and compare the outcome to the full calculation. 
In particular, we will assess the sensitivity of present DD
experiments to a set of operators up to dimension 7 describing dark
matter interactions with quarks $q$ and gluons
\begin{equation}\label{eq:lightDM:Lnf5}
{\cal L}_\chi=\sum_q \sum_{a,d}
\C_{a,q}^{(d)} {\cal Q}_{a,q}^{(d)}+\sum_{b,d}
\C_{b}^{(d)} {\cal Q}_{b}^{(d)}, 
\end{equation}
\noindent where the $\C_{a,q}^{(d)}$, $\C_{b}^{(d)}$ are dimensional
Wilson coefficients. The sums run over the dimensions of the
operators, $d=5,6,7$ and the operator labels, $a$ and $b$.  If not
specified otherwise, we conventionally fix the Wilson parameters at
the Electroweak (EW) scale, that we identify with the $Z$ boson mass.
The operators ${\cal Q}_{a,q}^{(d)}$, ${\cal Q}_{b}^{(d)}$ that we
will analyze are listed in
Eqs.(\ref{eq:dim5},\ref{eq:dim6},\ref{eq:dim7}), and are the same
analyzed in~\cite{bishara_2017}. Analyses on similar sets of
relativistic effective operators can also be found in~\cite{tait_2010,
  buckley_2013, desimone_2016, belyaev_reft}.

In particular, once, for each of the relativistic models we consider,
the NR Wilson coefficients $c_j^{\tau}$ at the nucleon scale are
obtained from the $\C_{a,q}^{(d)}$'s or $\C_{b}^{(d)}$'s, the expected
DD rates only depend on the non--relativistic response
functions~\cite{nreft_form_factors, peter_identify}.  In our analysis
we will follow closely Ref.~\cite{sogang_scaling_law_nr}, so that we
address the reader to that paper for the formulas that we use to
calculate the expected rates for WIMP--nucleus scattering.

In Ref.~\cite{sogang_scaling_law_nr} we found that 9 experiments out
of a total of 14 present Dark Matter searches can provide the most
stringent bound on some of the effective couplings for a given choice
of $(m_{\chi},c^n/c^p)$.  We include the same experiments in the
present analysis: XENON1T~\cite{xenon_2018},
CDMSlite~\cite{cdmslite_2017}, SuperCDMS~\cite{super_cdms_2017},
PICASSO~\cite{picasso}, PICO--60 (using a $CF_3I$ target
~\cite{pico60_2015} and a $C_3F_8$ one \cite{pico60_2019}), CRESST-II
\cite{cresst_II,cresst_II_ancillary}, DAMA (average count rate
\cite{damaz}), DarkSide--50 \cite{ds50}. The details of how each
experimental limit has been obtained can be found in the Appendix A of
Ref.~\cite{sogang_scaling_law_nr} with the exception of the PICO--60
result with a $C_3F_8$ target of Ref.~\cite{pico60} that was recently
updated with its final result~\cite{pico60_2019}. In particular,
in~\cite{pico60_2019} an additional exposure of 1404 kg days at
threshold 2.45 keVnr was included in the analysis, lower than that of
Ref.~\cite{pico60} where an exposure of 1167 kg days with threshold 3.3
keVnr was used. So, compared to Ref.~\cite{sogang_scaling_law_nr}, for
PICO--60 we have added the additional run at 2.45 keVnr and updated
the efficiency of both runs using the result from Fig.3
of~\cite{pico60_2019}.  We stress that in the literature only the
bounds from a few experiments (typically XENON1T and PICO--60) are
discussed, and only for a few of the effective models of
Eqs.(\ref{eq:dim5},\ref{eq:dim6},\ref{eq:dim7})~\cite{buckley_2013,
  deramo_2014, deramo_2016, bishara_2017, belyaev_2019,xenon100_nreft,
  cresst_nreft, super_cdms_nreft_2015}. So the second main goal of the
present paper is to focus on a systematic approach, including a number
of experiments and of effective couplings that both exceed for
completeness previous analyses.

The approach of the present analysis is complementary to that of
Ref.~\cite{sogang_scaling_law_nr}, but itself not devoid from
drawbacks.
In particular, the matching of the Wilson coefficients
$\C_{a,q}^{(d)}$'s of the WIMP--quark relativistic interaction into
the $c_j^{\tau}$'s of the NR WIMP--nucleon Hamiltonian is highly
degenerate, since, in principle, the relativistic effective theory
requires to fix one coupling for each quark flavor $q$, while the NR
theory contains only protons and neutrons. In other words, some
assumptions must be made on how the $\C_{a,q}^{(d)}$'s scale with the
flavor $q$.
A frequent approach in the literature is to parameterize the theory in
terms of a single coupling $\C_{a,q}^{(d)}$ common to all quarks~
\cite{Arina_2014,Belyaev_2018}, and in our analysis we will do the
same. However it is worth pointing out that this assumption would not
be applicable to the case of the supersymmetric neutralino, for which,
for instance, $\C_{4,q}^{(6)}$ scales as the $Z$--boson coupling in
the case of a Higgsino, or $\C_{5,q}^{(7)}$ depends on the mass and
the weak isospin of the quark for a Gaugino--Higgsino mixing. This has
important phenomenological consequences: for instance, the ratio of
the two vacuum expectation values in supersymmetry, traditionally
parameterized as $\tan\beta$, selects through the Yukawa couplings
whether the neutralino couples preferentially to up--type or
down--type quarks through Higgs exchange, and it is well known in the
literature that large and low $\tan\beta$ values imply very different
phenomenological scenarios.  Indeed, in the case of a generic scaling
of the WIMP--quark couplings the only possible way to obtain a
consistent limit without reanalyzing the experimental data is to
calculate the ratio $c^n/c^p$ from the $\C_{a,q}^{(d)}$'s and directly
use the NR bounds of \cite{sogang_scaling_law_nr}. The limit obtained
in this way is only valid if one NR coupling dominates the predicted
rate and there are no cancellations among the contributions of
different NR couplings.  So a specific goal of our analysis is also to
assess the validity of such a procedure and to discuss the impact of
such cancellations in the different relativistic models we consider.

The paper is organized as follows. In Section \ref{sec:rel_eft} we
list the relativistic Effective Field Theory (EFT) terms that we
consider in our analysis and we summarize how we calculate the NR
Wilson coefficients starting from each of them;
Section~\ref{sec:analysis} is devoted to our quantitative analysis,
where we will provide updated exclusion plots for each relativistic
model assuming a common coupling $\C_{a,q}^{(d)}$ for all quarks; in
Section \ref{sec:interferences} we discuss the impact of interferences
among different NR couplings, showing that in most cases only one
non--relativistic operator dominates the expected rate and the
bounds. We will provide our conclusions in Section
\ref{sec:conclusions}. Finally, in~\ref{app:program} we provide a
simple interpolation code written in Python that, based on the
conclusions of Section~\ref{sec:interferences}, allows to reproduce
most of the results of Section~\ref{sec:analysis} and to generalize
them to other choices of the $\C_{a,q}^{(d)}$ couplings assuming that
one non--relativistic operator dominates the expected rate.

\section{Relativistic effective models}
\label{sec:rel_eft}

In this Section we outline the procedure that we follow to obtained
the numerical results of Section~\ref{sec:analysis}. We use the code
DirectDM~\cite{bishara_2017, directdm} to calculate the
nonperturbative matching of the effective field theory describing dark
matter interactions with quarks and gluons at the EW scale to the
effective theory of nonrelativistic dark matter interacting with
nonrelativistic nucleons (alternative analyses based on chiral
effective field theory can be found for instance
in~\cite{matching_solon1,matching_solon2,chiral_eft,hoferichter_si}).
For this reason we follow closely the notation of
Ref.\cite{bishara_2017,reft_dim7} and consider the same relativistic
operators.

In particular, we consider the two dimension-five operators:
\begin{equation}
\label{eq:dim5}
{\cal Q}_{1}^{(5)} = \frac{e}{8 \pi^2} (\bar \chi \sigma^{\mu\nu}\chi)
 F_{\mu\nu} \,, \qquad {\cal Q}_2^{(5)} = \frac{e }{8 \pi^2} (\bar
\chi \sigma^{\mu\nu} i\gamma_5 \chi) F_{\mu\nu} \,,
\end{equation}
where $F_{\mu\nu}$ is the electromagnetic field strength tensor and
$\chi$ is the DM field, assumed here to be a Dirac particle. Such
operators correspond, respectively, to magnetic--dipole and
electric--dipole DM and imply a long--range
interaction~\cite{delnobile_2018} \footnote{The anapole coupling $(\bar
  \chi \gamma^{\mu}\gamma_5\chi) \partial^{\nu}F_{\mu\nu}$ leads
  instead to an effective contact interaction. A recent discussion is
  provided in~\cite{sogang_gondolo_anapole}.}.  The dimension-six
operators are
\begin{eqnarray}
{\cal Q}_{1,q}^{(6)} & =& (\bar \chi \gamma_\mu \chi) (\bar q \gamma^\mu q)\,,
 {\cal Q}_{2,q}^{(6)} = (\bar \chi\gamma_\mu\gamma_5 \chi)(\bar q \gamma^\mu q)\,, \nonumber
  \\ 
{\cal Q}_{3,q}^{(6)} & =& (\bar \chi \gamma_\mu \chi)(\bar q \gamma^\mu \gamma_5 q)\,,
   {\cal Q}_{4,q}^{(6)} = (\bar
\chi\gamma_\mu\gamma_5 \chi)(\bar q \gamma^\mu \gamma_5 q)\,,\label{eq:dim6}
\end{eqnarray}
and we also include the following dimension-seven operators:
namely: 
\begin{eqnarray}
{\cal Q}_1^{(7)} & =& \frac{\alpha_s}{12\pi} (\bar \chi \chi)
 G^{a\mu\nu}G_{\mu\nu}^a\,, 
  {\cal Q}_2^{(7)} = \frac{\alpha_s}{12\pi} (\bar \chi i\gamma_5 \chi) G^{a\mu\nu}G_{\mu\nu}^a\,,\nonumber
 \\
{\cal Q}_3^{(7)} & =& \frac{\alpha_s}{8\pi} (\bar \chi \chi) G^{a\mu\nu}\widetilde
 G_{\mu\nu}^a\,, 
 {\cal Q}_4^{(7)} = \frac{\alpha_s}{8\pi}
(\bar \chi i \gamma_5 \chi) G^{a\mu\nu}\widetilde G_{\mu\nu}^a \,, \nonumber
\\
{\cal Q}_{5,q}^{(7)} & =& m_q (\bar \chi \chi)( \bar q q)\,, 
{\cal
  Q}_{6,q}^{(7)} = m_q (\bar \chi i \gamma_5 \chi)( \bar q q)\,,\nonumber
  \\
{\cal Q}_{7,q}^{(7)} &=& m_q (\bar \chi \chi) (\bar q i \gamma_5 q)\,, 
{\cal Q}_{8,q}^{(7)}  = m_q (\bar \chi i \gamma_5 \chi)(\bar q i \gamma_5
q)\,, \nonumber  
 \\
{\cal Q}_{9,q}^{(7)} & =& m_q (\bar \chi \sigma^{\mu\nu} \chi) (\bar q \sigma_{\mu\nu} q)\,, 
{\cal Q}_{10,q}^{(7)}  = m_q (\bar \chi  i \sigma^{\mu\nu} \gamma_5 \chi)(\bar q \sigma_{\mu\nu}
q)\,. \label{eq:dim7} 
\end{eqnarray}
\noindent In the equations above $q=u,d,s$ denote the light quarks,
$G_{\mu\nu}^a$ is the QCD field strength tensor, while $\widetilde
G_{\mu\nu} = \frac{1}{2}\varepsilon_{\mu\nu\rho\sigma} G^{\rho\sigma}$
is its dual, and $a=1,\dots,8$ are the adjoint color indices. In the
following we will also assume that all the operators listed in
Eqs.(\ref{eq:dim5})--(\ref{eq:dim7}) conserve flavor.

A potentially sizeable mixing effect among the vector and
axial--vector currents of Eq.(\ref{eq:dim6}) is known to be induced by
the running of the couplings above the EW scale
~\cite{deramo_2014,deramo_2016}. In particular, this may induce a
quark vector coupling at the low scale relevant for DD even if the
effective theory contains only an axial coupling at the high scale,
changing dramatically the DD cross section scaling with the nuclear
target and the ensuing DD constrains. For this reason, in the case of
the operators of Eq. (\ref{eq:dim6}) with a vector--axial quark
current, besides the results valid for a given effective operator at
the EW scale we also show the corresponding ones when the same
operator is defined at the scale $\mu_{scale}$=2 TeV, using the code
runDM~\cite{rundm} to evaluate the running from $\mu_{scale}$ to
$m_Z$=91.1875 GeV. In order to do so we assume that the axial--vector
coupling is the same for all quarks at the high scale~\footnote{In
  particular, we assume the benchmark ``QuarksAxial" in~\cite{rundm},
  with a vanishing DM-Higgs coupling.}. We then use the output of
runDM as an input for DirectDM to perform the remaining running from
$m_Z$ to the nucleon scale, where the hadronization of the operators
${\cal Q}_{a,q}^{(d)}$ in
eqs. (\ref{eq:dim5},\ref{eq:dim6},\ref{eq:dim7}) leads at leading
order in the chiral expansion only to single-nucleon (N=p,n) currents,
i.e., schematically:

\begin{eqnarray}
  <N|\bar{q}\Gamma q|N>&=&\sum_{\Gamma'}\Omega_N^{\Gamma'} \bar{\Psi}_N \Gamma'
  \Psi_N,\nonumber\\ <N|G^{a\mu\nu}G_{a,\mu\nu}|N>&=&\Omega_N^{\prime}
  \bar{\Psi}_N \Psi_N,\nonumber\\ <N|
  G^{a\mu\nu}\tilde{G}_{a,\mu\nu}|N>&=&\Omega_N^{\prime\prime}
  \bar{\Psi}_N \gamma_5 \Psi_N,
  \label{eq:matching}
\end{eqnarray}

\noindent with $\Gamma$,
$\Gamma^{\prime}=1,\gamma^{\mu},\gamma^{\mu}\gamma_5,\gamma_5,\sigma^{\mu\nu}$
and $\Psi_N$ the nucleon field. Also for the quantities $\Omega$,
$\Omega^{\prime}$ and $\Omega^{\prime\prime}$ (which in general can
depend on external momenta) we rely on the output of DirectDM (see
appendix A of~\cite{bishara_2017}). In particular, the matching of the
axial-axial partonic level operator, as well as that of the coupling
between the DM particle to the QCD anomaly term leads to pion and eta
poles that can be numerically important, and that we include in our
analysis. Specifically~\cite{bishara_2017}:

\begin{eqnarray}
\label{axial:form:factor}
\langle N|\bar q \gamma^\mu \gamma_5 q|N\rangle&=&\bar{\Psi}_N \Big[F_A^{q/N}(q^2)\gamma^\mu\gamma_5+\frac{1}{2m_N}F_{P'}^{q/N}(q^2) \gamma_5 q^\mu\Big]\Psi_N\,,
\\
\label{pseudoscalar:form:factor}
\langle N| m_q \bar q  i \gamma_5 q|N\rangle&=& F_P^{q/N} (q^2)\, \bar{\Psi}_N i \gamma_5 \Psi_N\,,
\\
\label{CPodd:gluonic:form:factor}
\langle N| \frac{\alpha_s}{8\pi} G^{a\mu\nu}\tilde G^a_{\mu\nu}|N\rangle&=& F_{\tilde G}^{N} (q^2)\, \bar{\Psi}_N i \gamma_5 \Psi_N\, ,
\end{eqnarray}
\noindent with:

\begin{eqnarray}
\label{eq:F_PP'}
F_{P,P'}^{q/N}(q^2)&=&\frac{m_N^2}{m_\pi^2-q^2} a_{\pi}^{q/N}+\frac{m_N^2}{m_\eta^2-q^2} a_{\eta}^{q/N}+b^{q/N}
,
\\
\label{eq:F_tildeG}
F_{\tilde G}^{N}(q^2)&=&
\frac{q^2}{m_\pi^2-q^2} a_{\tilde G,\pi}^{N}+\frac{q^2}{m_\eta^2-q^2} a_{\tilde G,\eta}^{N}+b_{\tilde G}^{N},
\end{eqnarray}

\noindent where we use DirectDM and runDM when applicable to calculate
the coefficients $a$ and $b$ from the high--energy couplings, and
$q^2$ represents here the squared four--momentum transfer. Finally,
taking the non--relativistic limit, we obtain the coefficients
$c_{i}^{\tau}$ of the effective Hamiltonian of Eq.~(\ref{eq:H}), which
turn out to be proportional to the initial relativistic dimensional
coupling $\C_{a,q}^{(d)}$, and, in general, depend on the WIMP mass
$m_{\chi}$ and on the exchanged momentum $q$ (the latter dependence
both through the poles of Eqs.(\ref{eq:F_PP'},\ref{eq:F_tildeG}) and
because of the photon propagator induced by the dimension--5 magnetic and
electric dipole operators of Eq. (\ref{eq:dim5})).

For the details of the expression to calculate the expected rate in a
DD experiment we refer, for instance, to Section 2 of
\cite{sogang_scaling_law_nr}.  In particular, the differential cross
section is proportional to the squared amplitude:

\be
\frac{d\sigma_T}{d E_R}=\frac{2 m_T}{4\pi v_T^2}\left [ \frac{1}{2 j_{\chi}+1} \frac{1}{2 j_{T}+1}|\mathcal{M}_T|^2 \right ],
\label{eq:dsigma_de}
\ee

\noindent with $v_T\equiv|\vec{v}_T|$ the WIMP speed in the reference
frame of the nuclear center of mass, $m_T$ the nuclear mass, $j_T$,
$j_\chi$ are the spins of target nucleus and WIMP, and \cite{haxton2}:

\begin{equation}
  \frac{1}{2 j_{\chi}+1} \frac{1}{2 j_{T}+1}|\mathcal{M}_T|^2=
  \frac{4\pi}{2 j_{T}+1} \sum_{\tau=0,1}\sum_{\tau^{\prime}=0,1}\sum_{k} R_k^{\tau\tau^{\prime}}\left [c^{\tau}_i,c_j^{\tau^{\prime}},(v^{\perp}_T)^2,\frac{q^2}{m_N^2}\right ] W_{T k}^{\tau\tau^{\prime}}(y).
\label{eq:squared_amplitude}
\end{equation}

\noindent In the above expression the squared amplitude
$|\mathcal{M}_T|^2$ is summed over initial and final spins, the
$R_k^{\tau\tau^{\prime}}$'s are WIMP response functions which depend
on the couplings $c^{\tau}_j$ as well as the transferred momentum
$\vec{q}$, while:

\begin{equation}
(v^{\perp}_T)^2=v^2_T-v_{min}^2,
\label{eq:v_perp}
\end{equation}
and:
\begin{equation}
v_{min}^2=\frac{q^2}{4 \mu_{T}^2}=\frac{m_T E_R}{2 \mu_{T}^2},
\label{eq:vmin}
\end{equation}

\noindent represents the minimal incoming WIMP speed required to
impart the nuclear recoil energy $E_R$. Moreover, in equation
(\ref{eq:squared_amplitude}) the $W^{\tau\tau^{\prime}}_{T k}(y)$'s
are nuclear response functions and the index $k$ represents different
effective nuclear operators, which, under the assumption that the
nuclear ground state is an approximate eigenstate of $P$ and $CP$, can
be at most eight: following the notation in \cite{haxton1,haxton2},
$k$=$M$, $\Phi^{\prime\prime}$, $\Phi^{\prime\prime}M$,
$\tilde{\Phi}^{\prime}$, $\Sigma^{\prime\prime}$, $\Sigma^{\prime}$,
$\Delta$, $\Delta\Sigma^{\prime}$. The $W^{\tau\tau^{\prime}}_{T
  k}(y)$'s are function of $y\equiv (qb/2)^2$, where $b$ is the size
of the nucleus. For the target nuclei $T$ used in most direct
detection experiments the functions $W^{\tau\tau^{\prime}}_{T k}(y)$,
calculated using nuclear shell models, have been provided in
Refs.~\cite{haxton2,catena}. Details about the definitions of both the
functions $R_k^{\tau\tau^{\prime}}$'s and $W^{\tau\tau^{\prime}}_{T
  k}(y)$'s can be found in \cite{haxton2}. In particular, using the
decomposition:

\be
R_k^{\tau\tau^{\prime}}=R_{0k}^{\tau\tau^{\prime}}+R_{1k}^{\tau\tau^{\prime}}(v^{\perp}_T)^2=R_{0k}^{\tau\tau^{\prime}}+R_{1k}^{\tau\tau^{\prime}}\left ( v_T^2-v_{min}^2 \right ),
\label{eq:r_decomposition}
\ee

\noindent the correspondence between each term of the NR effective
interaction in~(\ref{eq:H}) and the $W^{\tau\tau^{\prime}}_{T k}(y)$
nuclear response functions is summarized in Table
\ref{table:eft_summary}. Notice that $W_M$ corresponds to the
standard SI interaction, while $W_{\Sigma^{\prime\prime}}+W_{\Sigma^{\prime}}$
(with $W_{\Sigma^{\prime}}\simeq 2 W_{\Sigma^{\prime\prime}}$) to the standard
SD one.

\begin{table}[t]
\begin{center}
{\begin{tabular}{@{}|c|c|c|c|c|c|@{}}
\hline
coupling  &  $R^{\tau \tau^{\prime}}_{0k}$  & $R^{\tau \tau^{\prime}}_{1k}$ & coupling  &  $R^{\tau \tau^{\prime}}_{0k}$  & $R^{\tau \tau^{\prime}}_{1k}$ \\
\hline
$1$  &   $M(q^0)$ & - & $3$  &   $\Phi^{\prime\prime}(q^4)$  & $\Sigma^{\prime}(q^2)$\\
$4$  & $\Sigma^{\prime\prime}(q^0)$,$\Sigma^{\prime}(q^0)$   & - & $5$  &   $\Delta(q^4)$  & $M(q^2)$\\
$6$  & $\Sigma^{\prime\prime}(q^4)$ & - & $7$  &  -  & $\Sigma^{\prime}(q^0)$\\
$8$  & $\Delta(q^2)$ & $M(q^0)$ & $9$  &  $\Sigma^{\prime}(q^2)$  & - \\
$10$  & $\Sigma^{\prime\prime}(q^2)$ & - & $11$  &  $M(q^2)$  & - \\
$12$  & $\Phi^{\prime\prime}(q^2)$,$\tilde{\Phi}^{\prime}(q^2)$ & $\Sigma^{\prime\prime}(q^0)$,$\Sigma^{\prime}(q^0)$ & $13$  & $\tilde{\Phi}^{\prime}(q^4)$  & $\Sigma^{\prime\prime}(q^2)$ \\
$14$  & - & $\Sigma^{\prime}(q^2)$ & $15$  & $\Phi^{\prime\prime}(q^6)$  & $\Sigma^{\prime}(q^4)$ \\
\hline
\end{tabular}}
\caption{Nuclear response functions corresponding to each coupling,
  for the velocity--independent and the velocity--dependent components
  parts of the WIMP response function, decomposed as in
  Eq.(\ref{eq:r_decomposition}).  In parenthesis is the power of $q$ in
  the WIMP response function.
  \label{table:eft_summary}}
\end{center}
\end{table}

Finally, for the WIMP local density we take $\rho_{loc}$=0.3
GeV/cm$^3$ and for the velocity distribution we assume a standard
isotropic Maxwellian at rest in the Galactic rest frame boosted to the
Lab frame by the velocity of the Sun, $v_{\odot}$=232 km/s, with
root--mean--square velocity $v_{rms}$=270 km/s and truncated at the
escape velocity $u_{esc}$=550 km/s.
\section{Analysis}
\label{sec:analysis}

In this Section for each of the models $\Q^{(d)}_{a,q}$,
$\Q^{(d)}_{b}$ listed in Eqs.~(\ref{eq:dim5}--\ref{eq:dim7}) we show
the present constraints on the correspondent dimensional coupling
$\C^{(d)}_{a,q}$ (assumed to be the same for all flavors) and
$\C^{(d)}_{b}$ from the list of experiments summarized in the
Introduction (XENON1T~\cite{xenon_2018},
CDMSlite~\cite{cdmslite_2017}, SuperCDMS~\cite{super_cdms_2017},
PICASSO~\cite{picasso}, PICO--60 (using a $CF_3I$ target
~\cite{pico60_2015} and a $C_3F_8$ one \cite{pico60_2019}), CRESST-II
\cite{cresst_II,cresst_II_ancillary}, DAMA (average count rate)
\cite{damaz}, DarkSide--50 \cite{ds50}) in terms of lower bounds on
the effective scale $\tilde{\Lambda}$ defined through:

\begin{equation}
  \C^{(d)}_{a,q},\C^{(d)}_{b}\equiv \frac{1}{\tilde{\Lambda}^{d-4}}.
  \label{eq:lambda}
\end{equation}

\noindent As a default choice in all cases we fix $\C^{(d)}_{a,q}$,
$\C^{(d)}_{b}$ at the EW scale, identified as the $Z$--boson mass,
$\mu_{scale}$= $m_Z$. Only for the 6--dimensional interaction terms
$\Q^{(6)}_{3,q}$ and $\Q^{(6)}_{4,q}$ we also show in
Fig.~\ref{fig:q6_34_2TeV_lambda} the result obtained when the
$\C^{(d)}_{a,q}$ coupling is fixed at the scale $\mu_{scale}$= 2 TeV
and run down to the EW scale using runDM and assuming that the
axial--vector coupling is the same for all quarks at the high
scale. One can notice that in the case of $\Q^{(6)}_{3,q}$ passing
from $\mu_{scale}$=$m_Z$ to $\mu_{scale}$=2 TeV the experimental bound
is strengthened by more than two orders of magnitude. This is a
well--known effect~\cite{deramo_2014,deramo_2016} due to the mixing
between $\Q^{(6)}_{3,q}$ and $\Q^{(6)}_{1,q}$ induced by the running
from 2 TeV to $m_Z$. In particular, without such mixing
$\Q^{(6)}_{3,q}$ gives rise to two NR operators that have both a
spin--dependent type scaling with the target, ${\cal O}_9$ and ${\cal
  O}_7$, the latter also velocity suppressed
\cite{sogang_scaling_law_nr}, while the mixing due to running induces
a $\Q^{(6)}_{1,q}$ component leading to the SI ${\cal O}_1$ operator
(see Tables~\ref{table:eft_summary} and~\ref{table:interferences})
that overwhelms the other contributions in spite of the
loop--suppressed Wilson coefficient. Such effect is also present for
the $\Q^{(6)}_{4,q}$ operator due to the mixing with $\Q^{(6)}_{2,q}$,
although in this case the effect on the exclusion plot is less
sizable. It is worth pointing out here that the mixing between vector
and vector--axial currents is driven by the coupling between the DM
particle and the quarks of the third family, so it is not present if
the latter is assumed to vanish. In such case the results of
Fig.~\ref{fig:q6_34_2TeV_lambda} would coincide to those of
Fig.~\ref{fig:q6_34_lambda}.

\begin{figure}
\begin{center}
  \includegraphics[width=0.49\columnwidth]{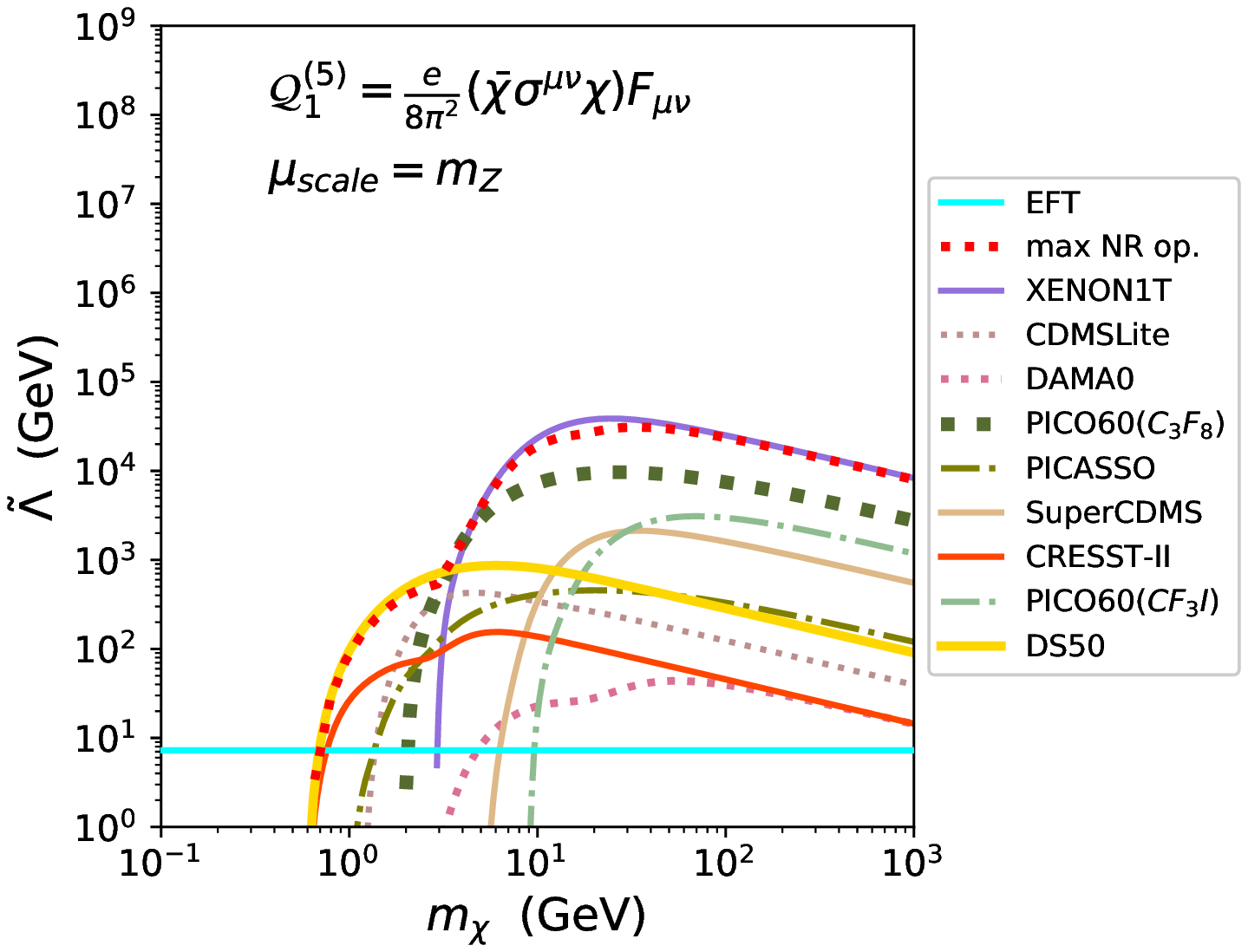}
  \includegraphics[width=0.49\columnwidth]{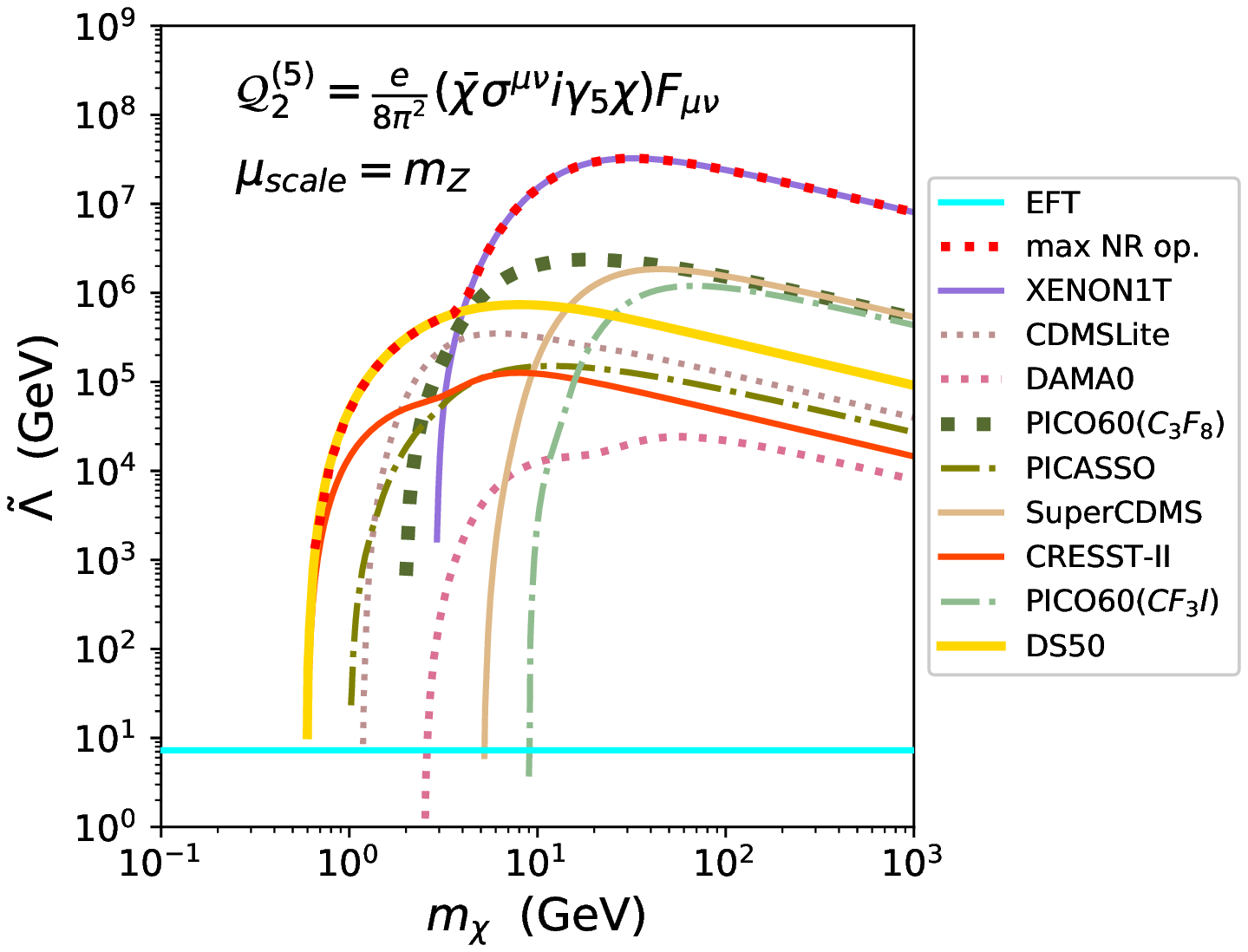}
\end{center}
\caption{Lower bound on the effective scale $\tilde{\Lambda}$ defined
  in Eq. (\ref{eq:lambda}) for the operators $\Q^{(5)}_{1,q}$ {\bf
    (left)} and $\Q^{(5)}_{2,q}$ {\bf (right)}. In both cases the
  dimensional couplings $\C^{(5)}_{1,q}$ and $\C^{(5)}_{2,q}$ are
  fixed at the EW scale $\mu_{scale}$=$m_Z$. In the region below the
  solid cyan line the limits are inconsistent with the validity of
  the EFT based on the simple criterion introduced in
  Section~\ref{sec:analysis}.
\label{fig:q5_lambda}}
\end{figure}

\begin{figure}
\begin{center}
  \includegraphics[width=0.49\columnwidth]{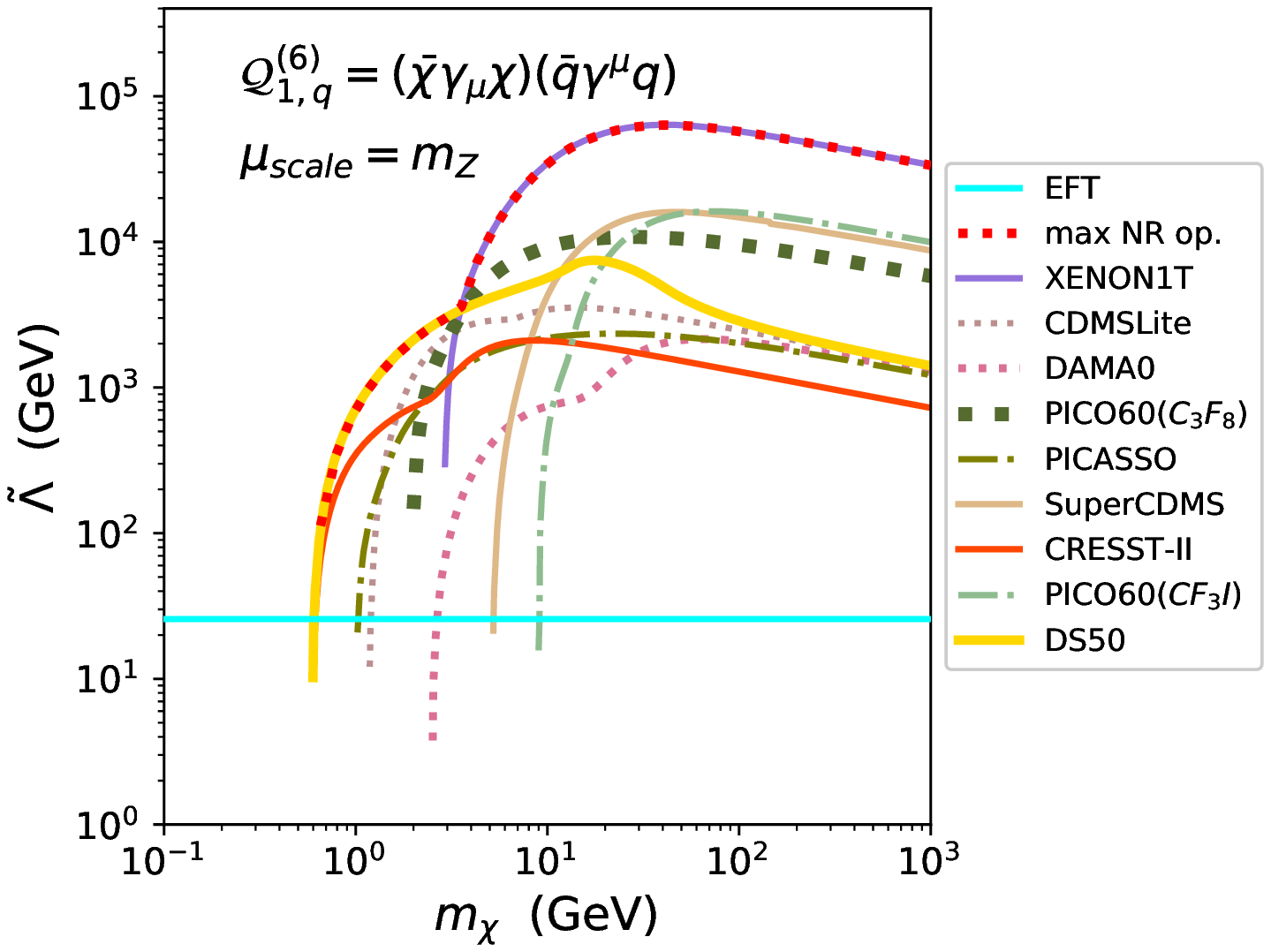}
  \includegraphics[width=0.49\columnwidth]{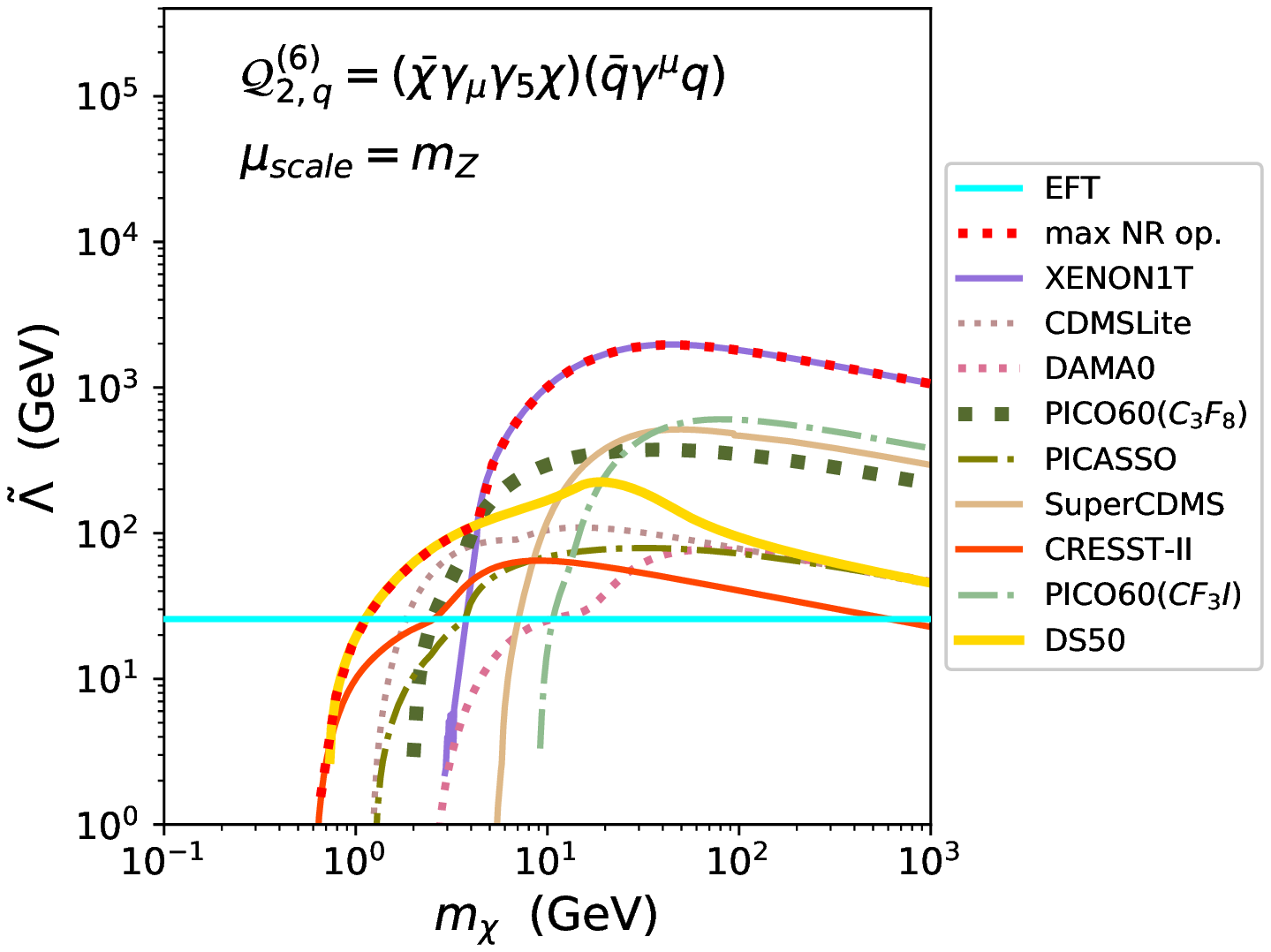}
\end{center}
\caption{The same as in Fig.~\ref{fig:q5_lambda} for $\Q^{(6)}_{1,q}$
  {\bf (left)} and $\Q^{(6)}_{2,q}$ {\bf (right)}.
\label{fig:q6_12_lambda}}
\end{figure}

\begin{figure}
\begin{center}
  \includegraphics[width=0.49\columnwidth]{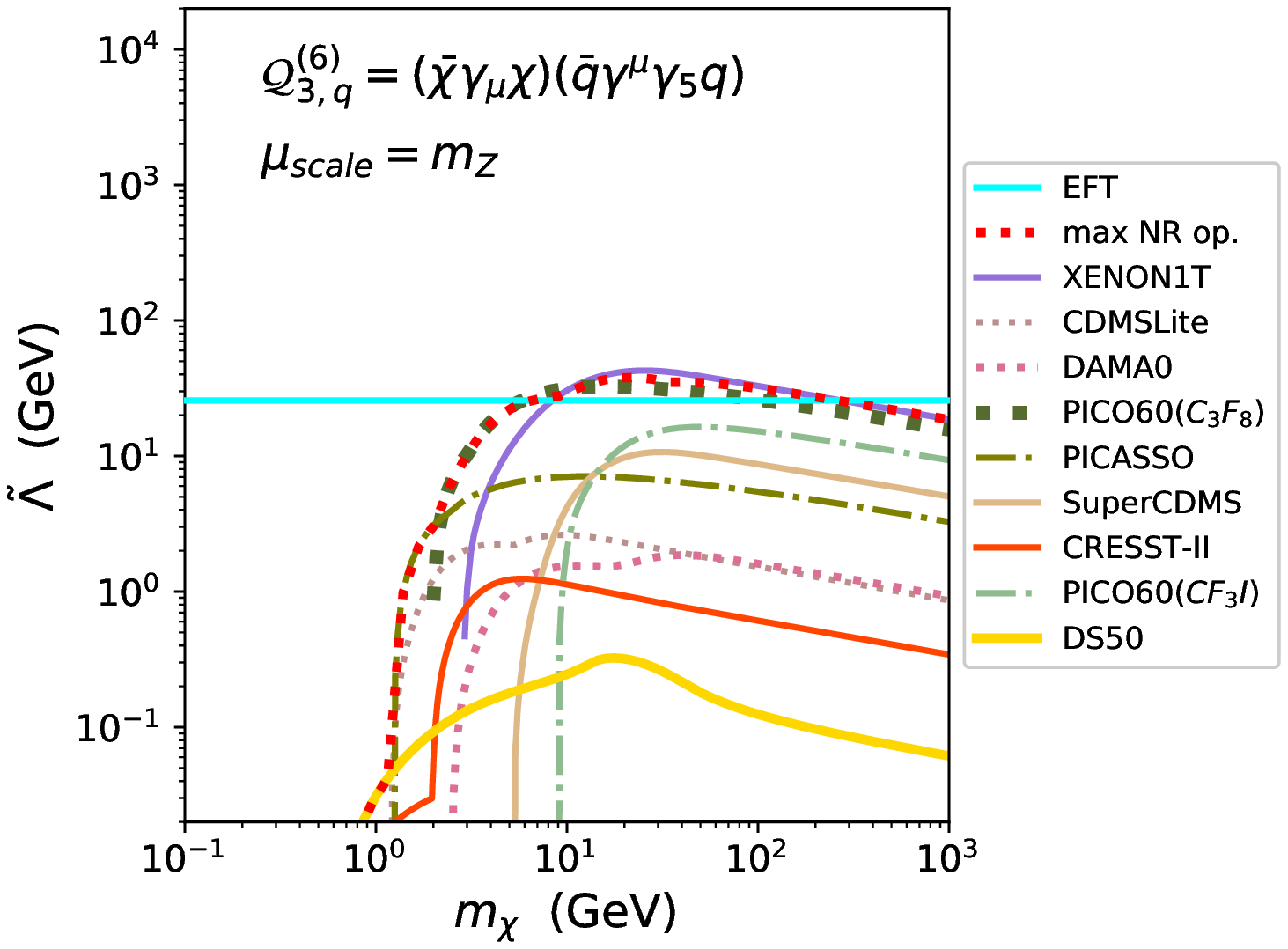}
  \includegraphics[width=0.49\columnwidth]{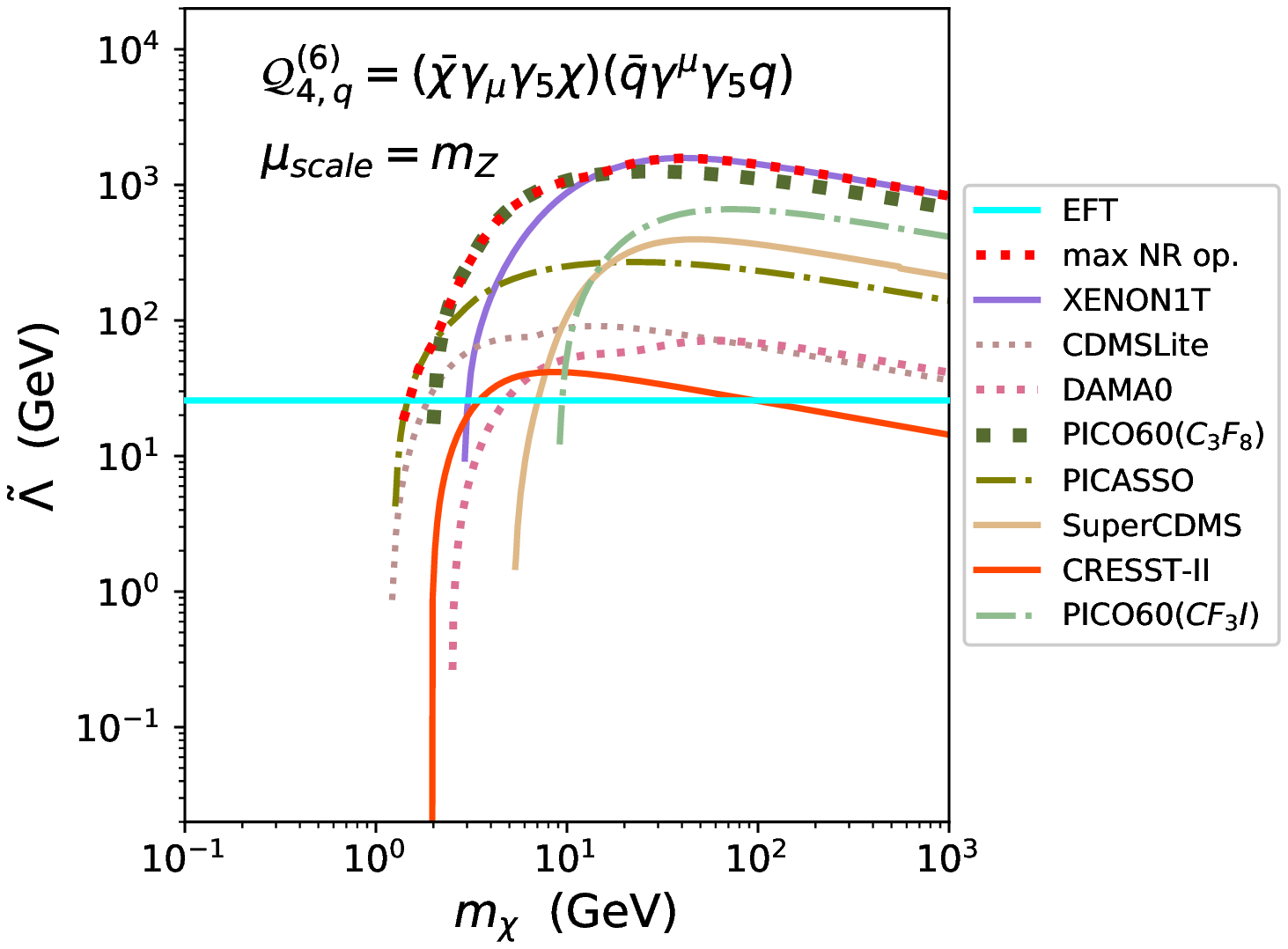}
\end{center}
\caption{The same as in Fig.~\ref{fig:q6_12_lambda} for $\Q^{(6)}_{3,q}$
  {\bf (left)} and $\Q^{(6)}_{4,q}$ {\bf (right)}.}
\label{fig:q6_34_lambda}
\end{figure}

\begin{figure}
\begin{center}
  \includegraphics[width=0.49\columnwidth]{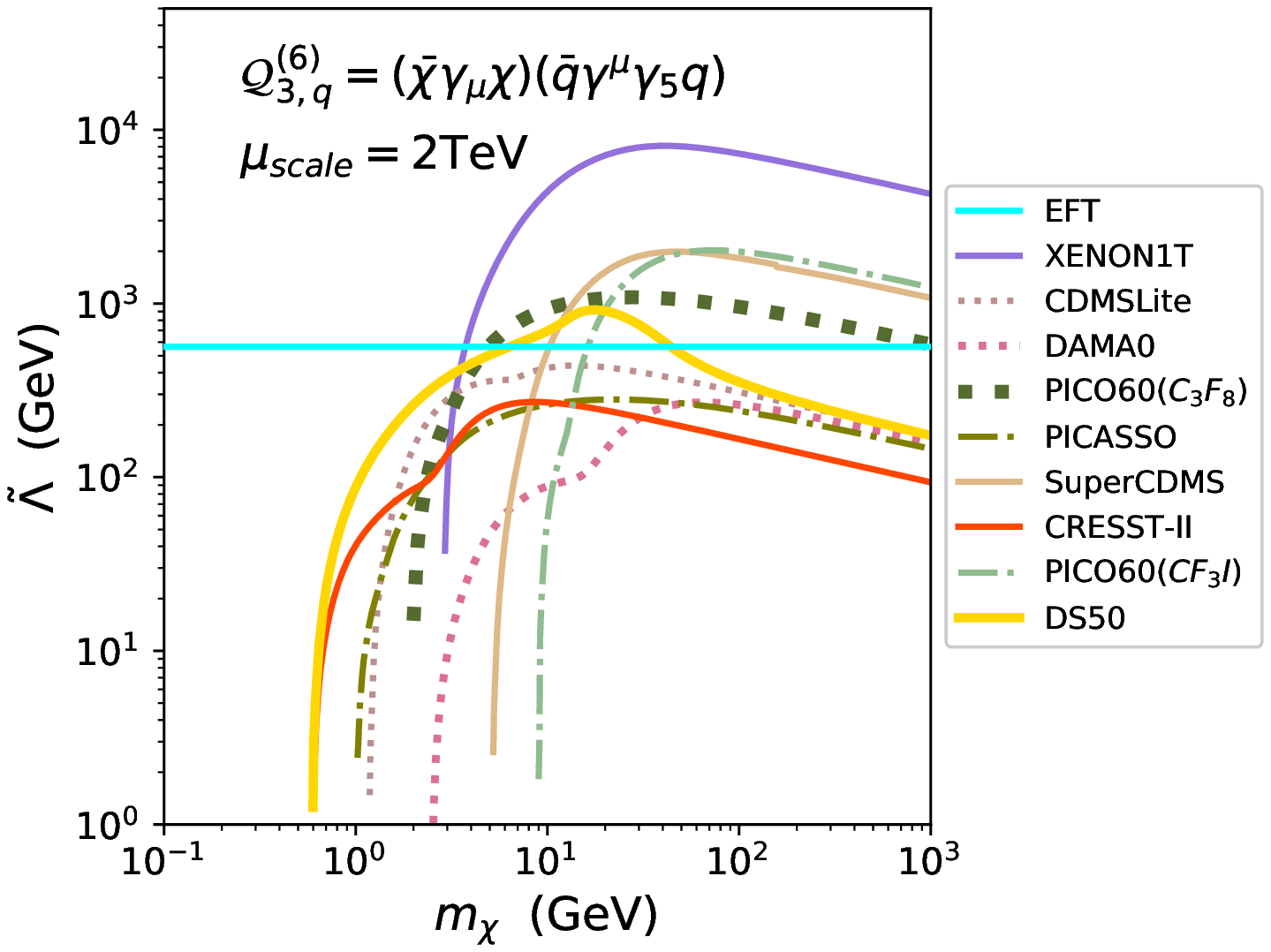}
  \includegraphics[width=0.49\columnwidth]{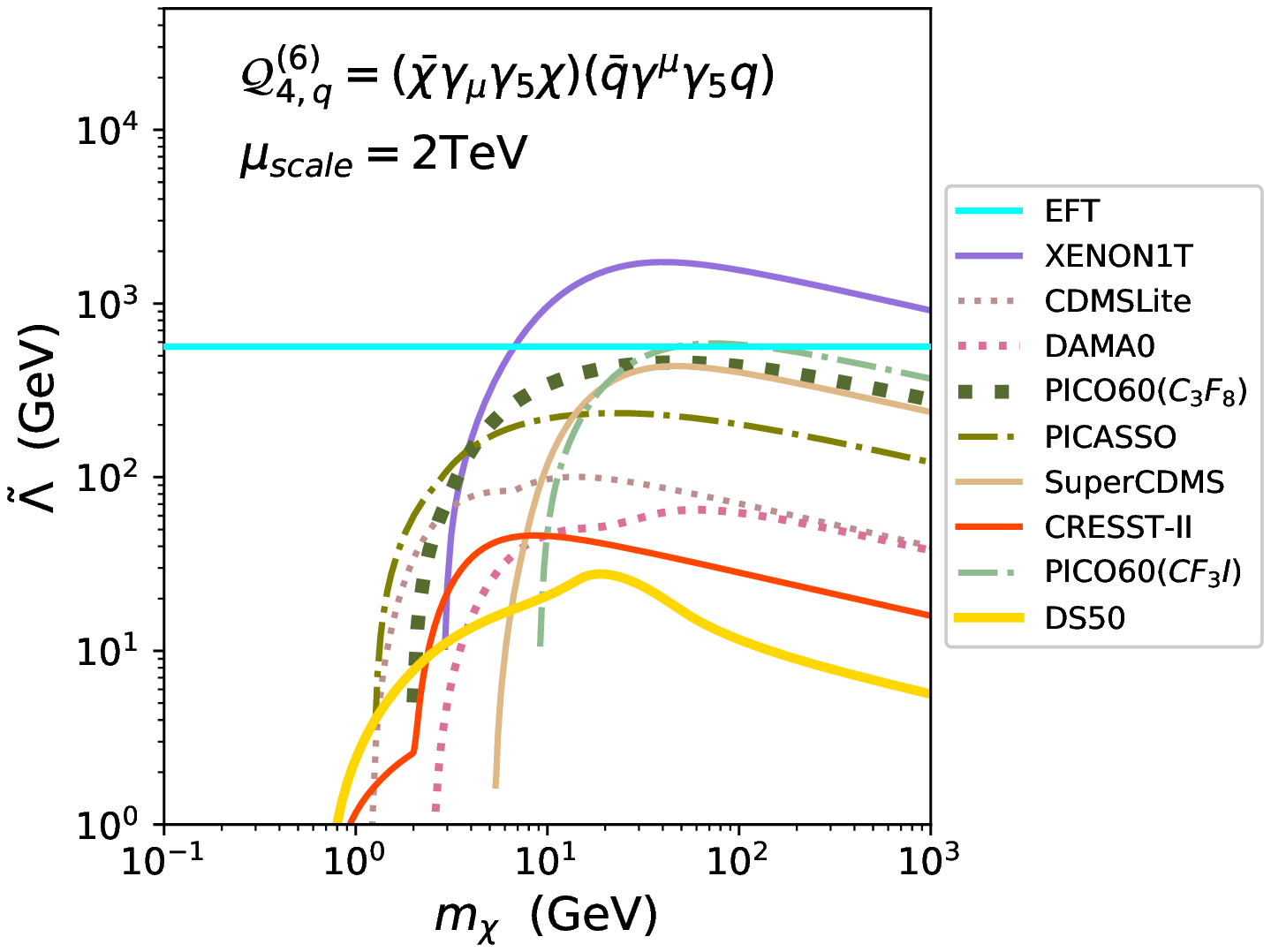}
\end{center}
\caption{The same as in Fig.~\ref{fig:q6_34_lambda} for
  $\mu_{scale}$=2 TeV.}
\label{fig:q6_34_2TeV_lambda}
\end{figure}

\begin{figure}
\begin{center}
  \includegraphics[width=0.49\columnwidth]{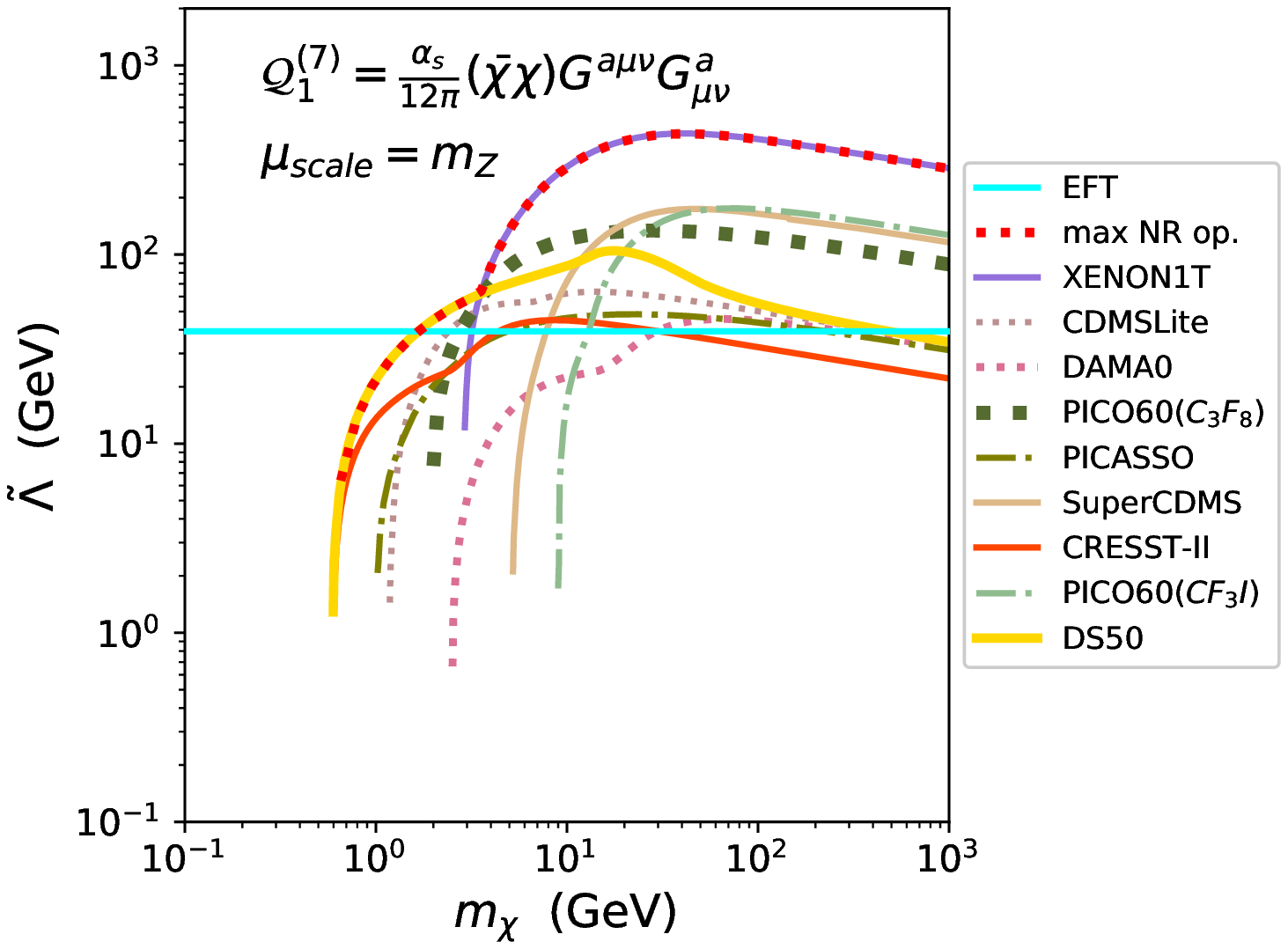}
  \includegraphics[width=0.49\columnwidth]{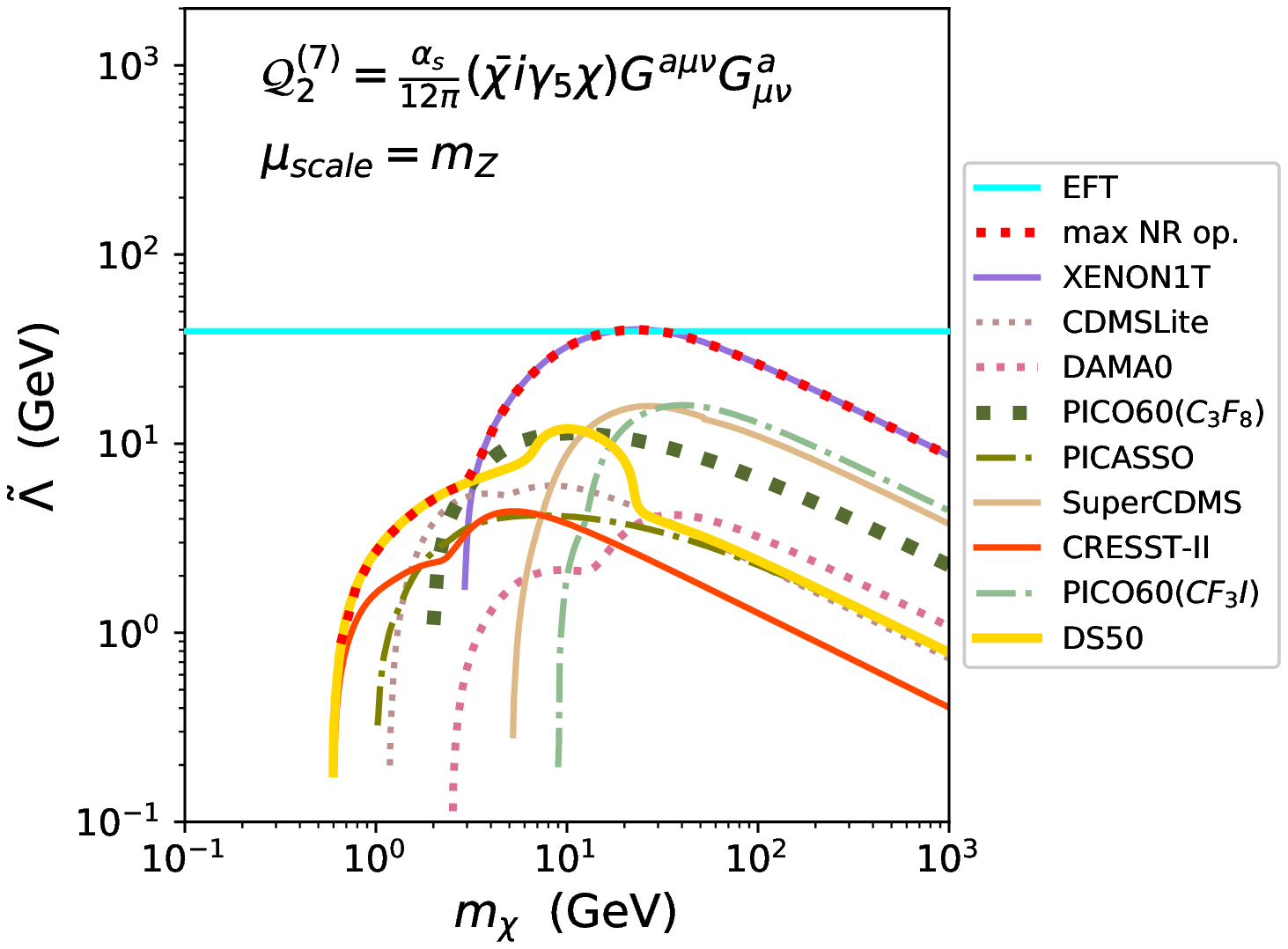}  
\end{center}
\caption{The same as in Fig.~\ref{fig:q6_12_lambda} for $\Q^{(7)}_{1}$
  {\bf (left)} and $\Q^{(7)}_{2}$ {\bf (right)}.}
\label{fig:q7_12_lambda}
\end{figure}

\begin{figure}
\begin{center}
  \includegraphics[width=0.49\columnwidth]{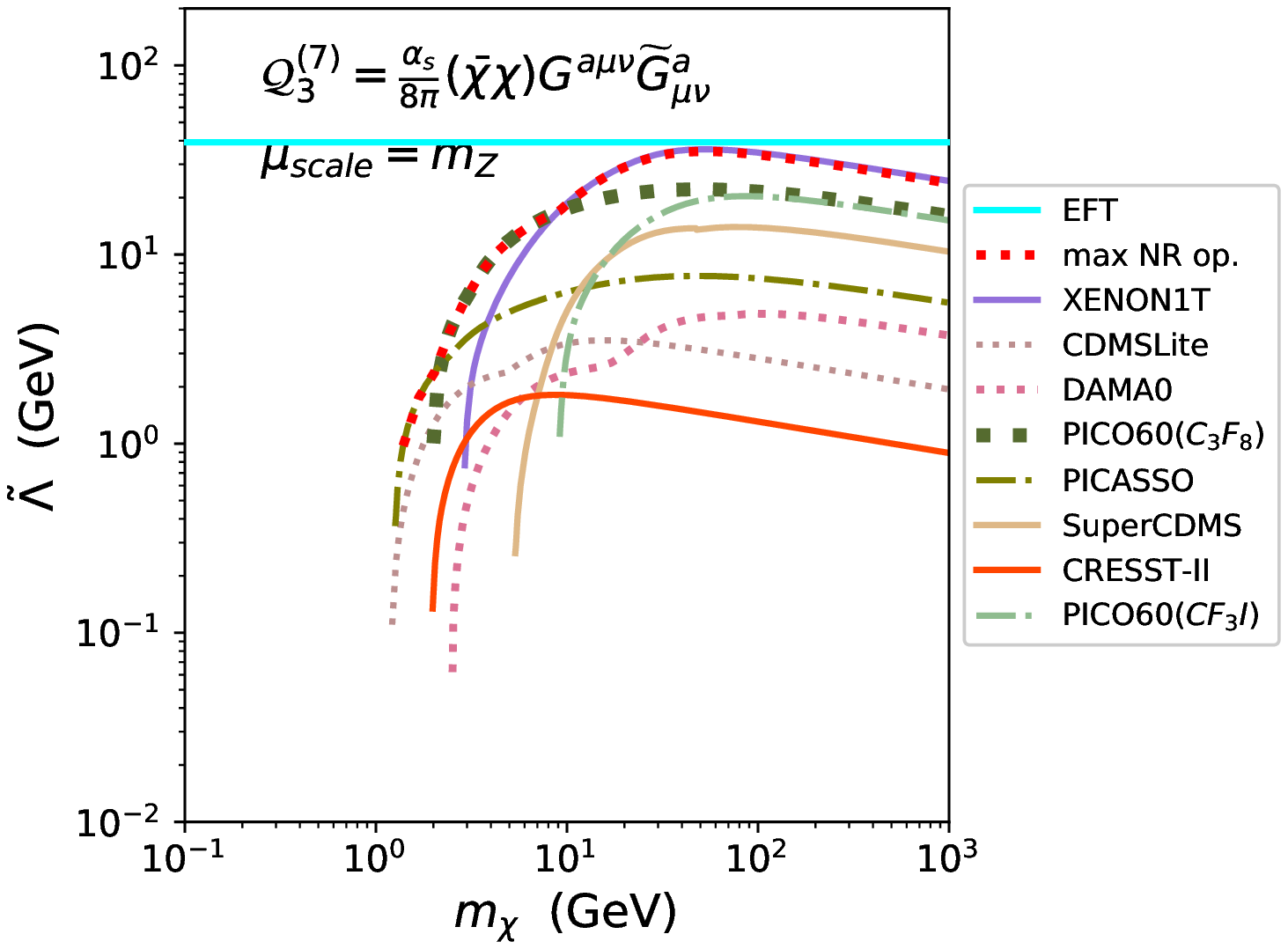}
  \includegraphics[width=0.49\columnwidth]{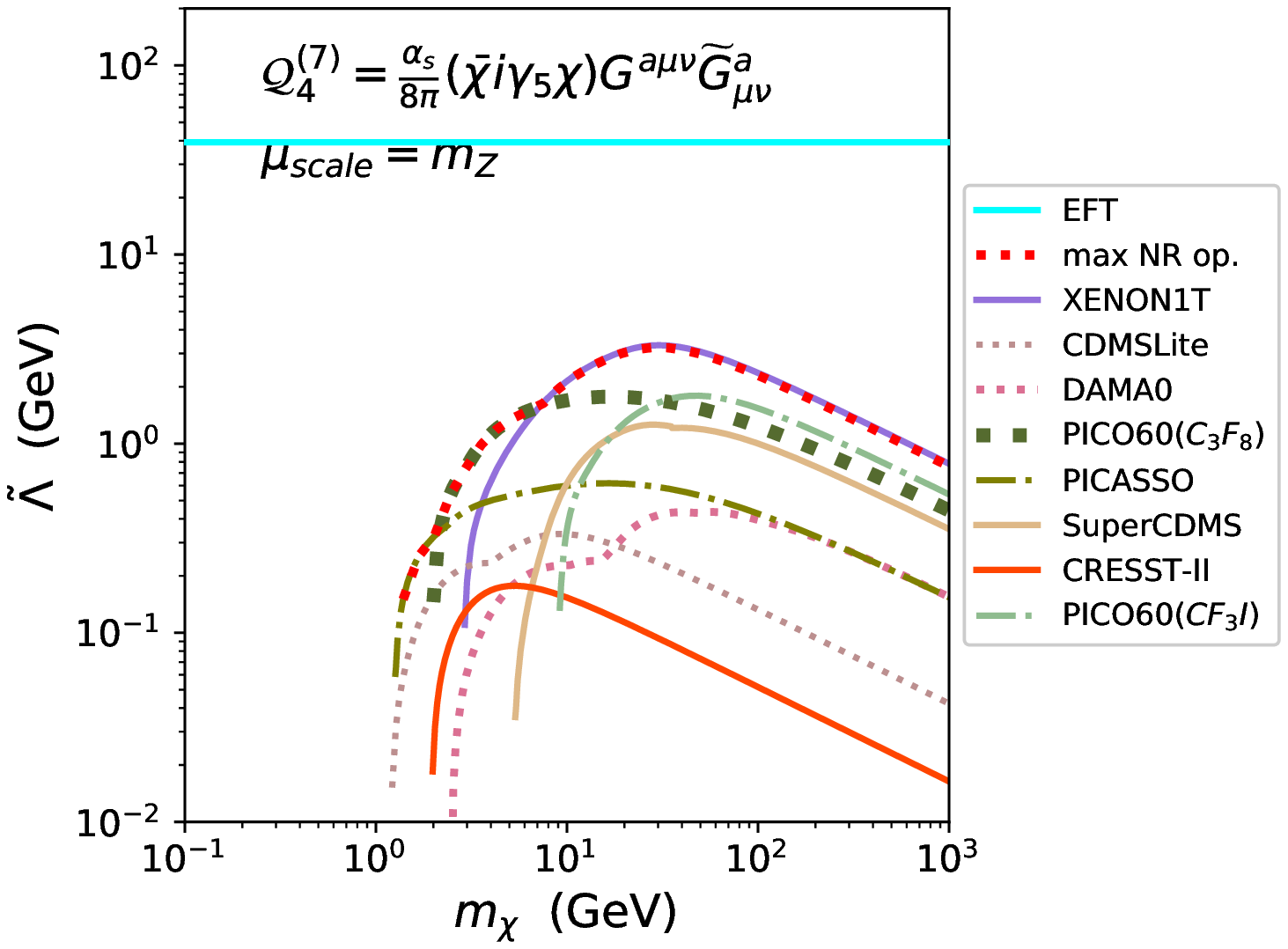}  
\end{center}
\caption{The same as in Fig.~\ref{fig:q6_12_lambda} for $\Q^{(7)}_{3}$
  {\bf (left)} and $\Q^{(7)}_{4}$ {\bf (right)}.}
\label{fig:q7_34_lambda}
\end{figure}

\begin{figure}
\begin{center}
  \includegraphics[width=0.49\columnwidth]{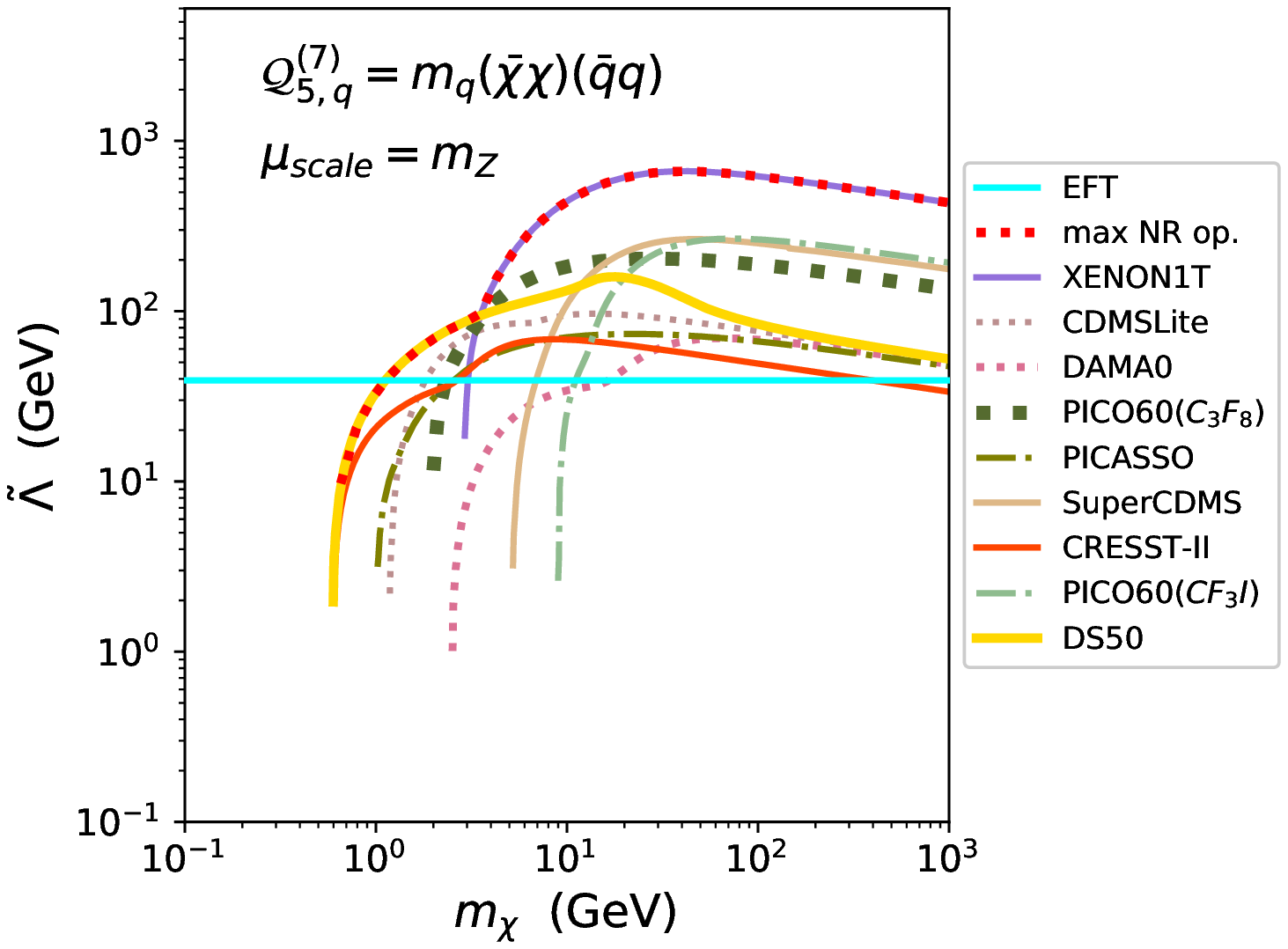}
  \includegraphics[width=0.49\columnwidth]{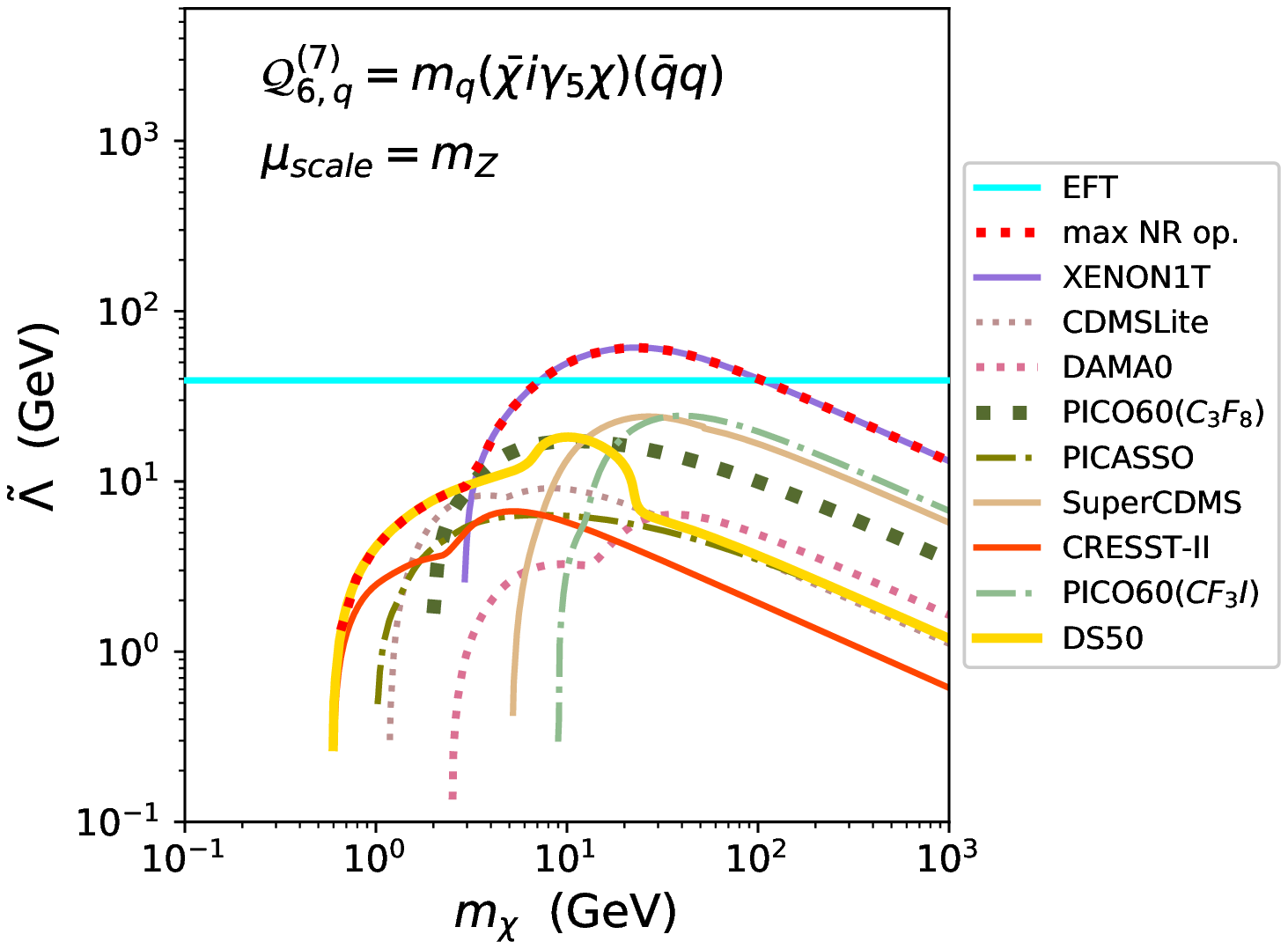}  
\end{center}
\caption{The same as in Fig.~\ref{fig:q6_12_lambda} for $\Q^{(7)}_{5}$
  {\bf (left)} and $\Q^{(7)}_{6}$ {\bf (right)}.}
\label{fig:q7_56_lambda}
\end{figure}

\begin{figure}
\begin{center}
  \includegraphics[width=0.49\columnwidth]{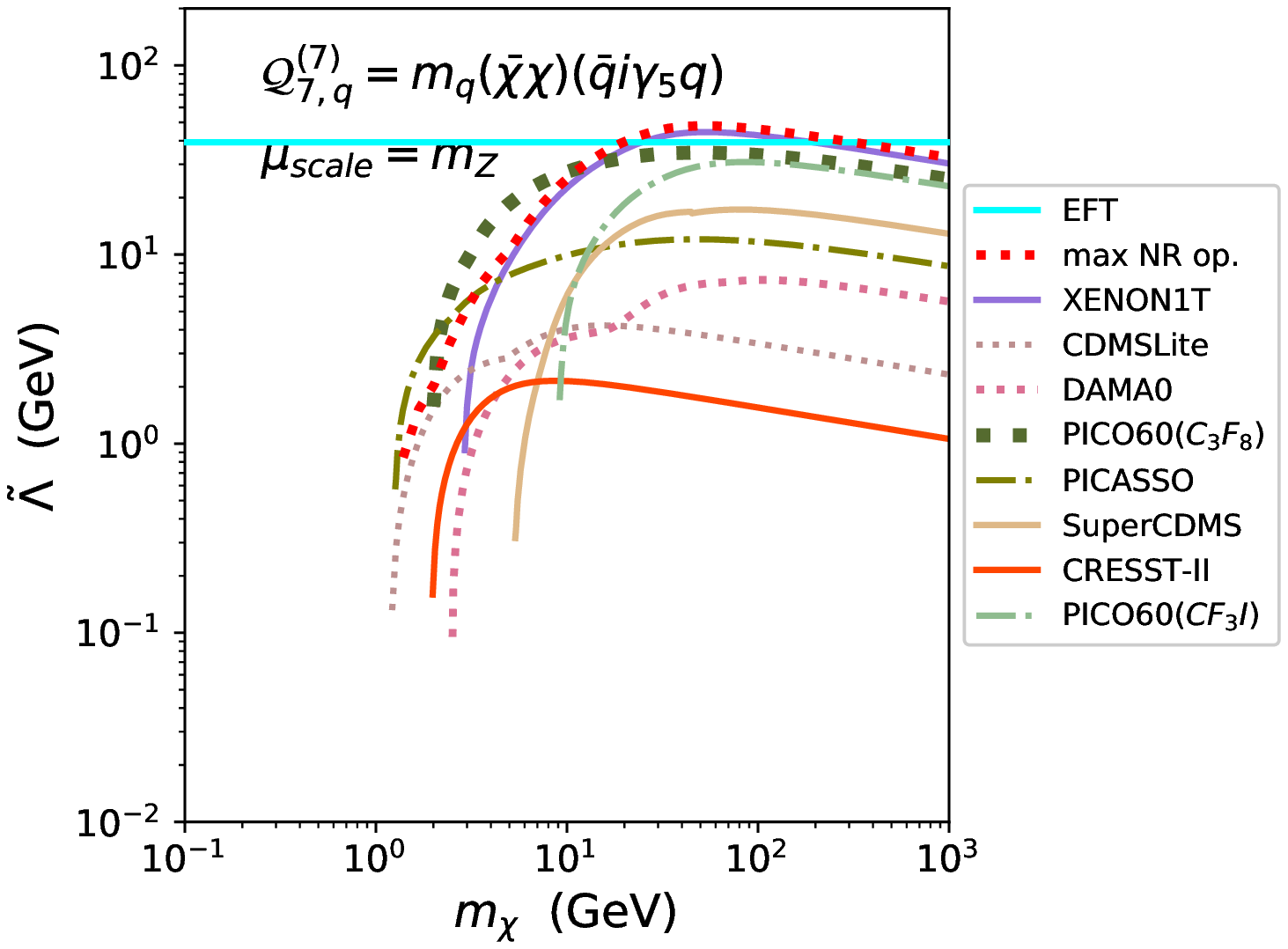}
  \includegraphics[width=0.49\columnwidth]{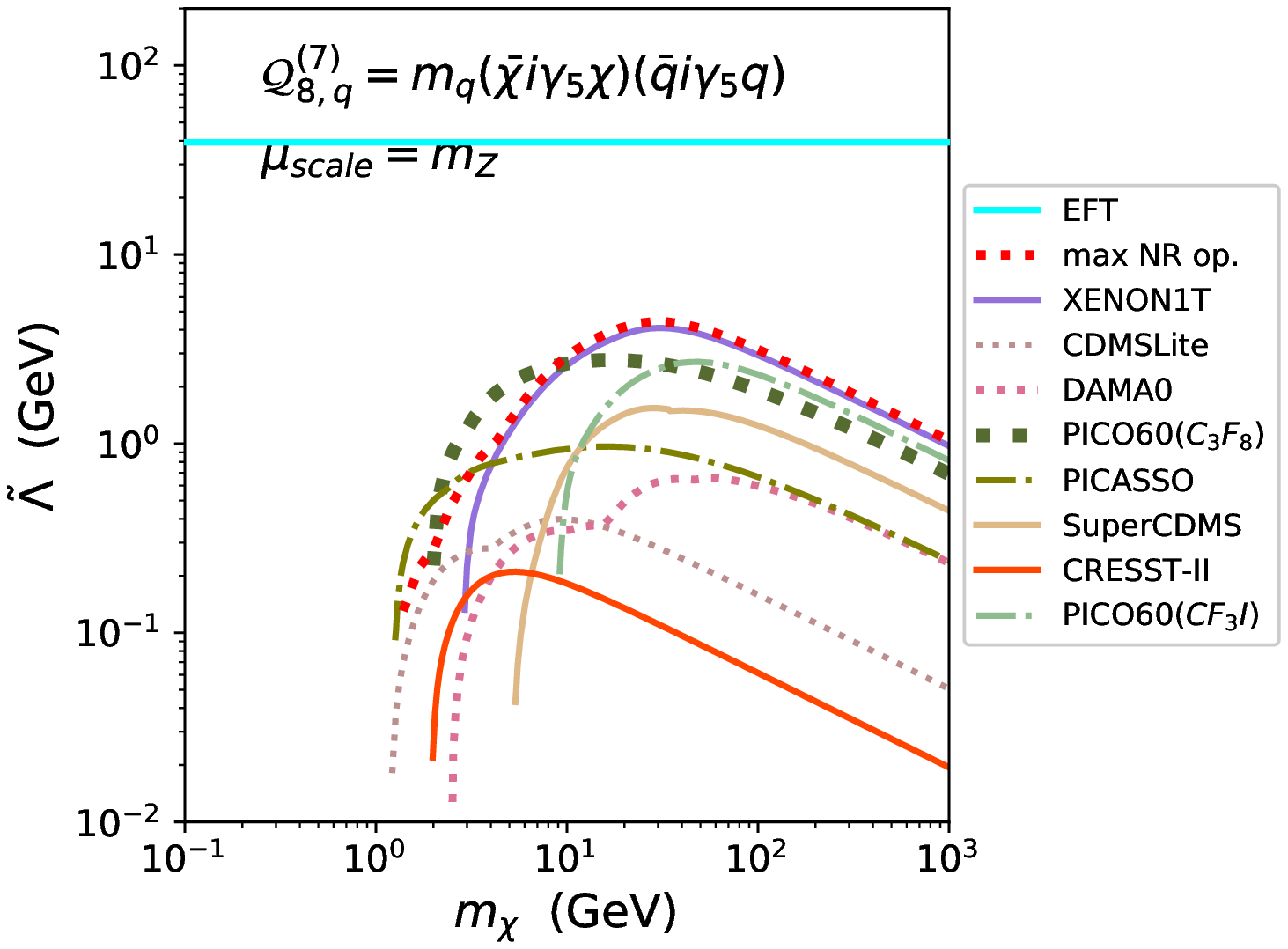}  
\end{center}
\caption{The same as in Fig.~\ref{fig:q6_12_lambda} for $\Q^{(7)}_{7}$
  {\bf (left)} and $\Q^{(7)}_{8}$ {\bf (right)}.}
\label{fig:q7_78_lambda}
\end{figure}

\begin{figure}
\begin{center}
  \includegraphics[width=0.49\columnwidth]{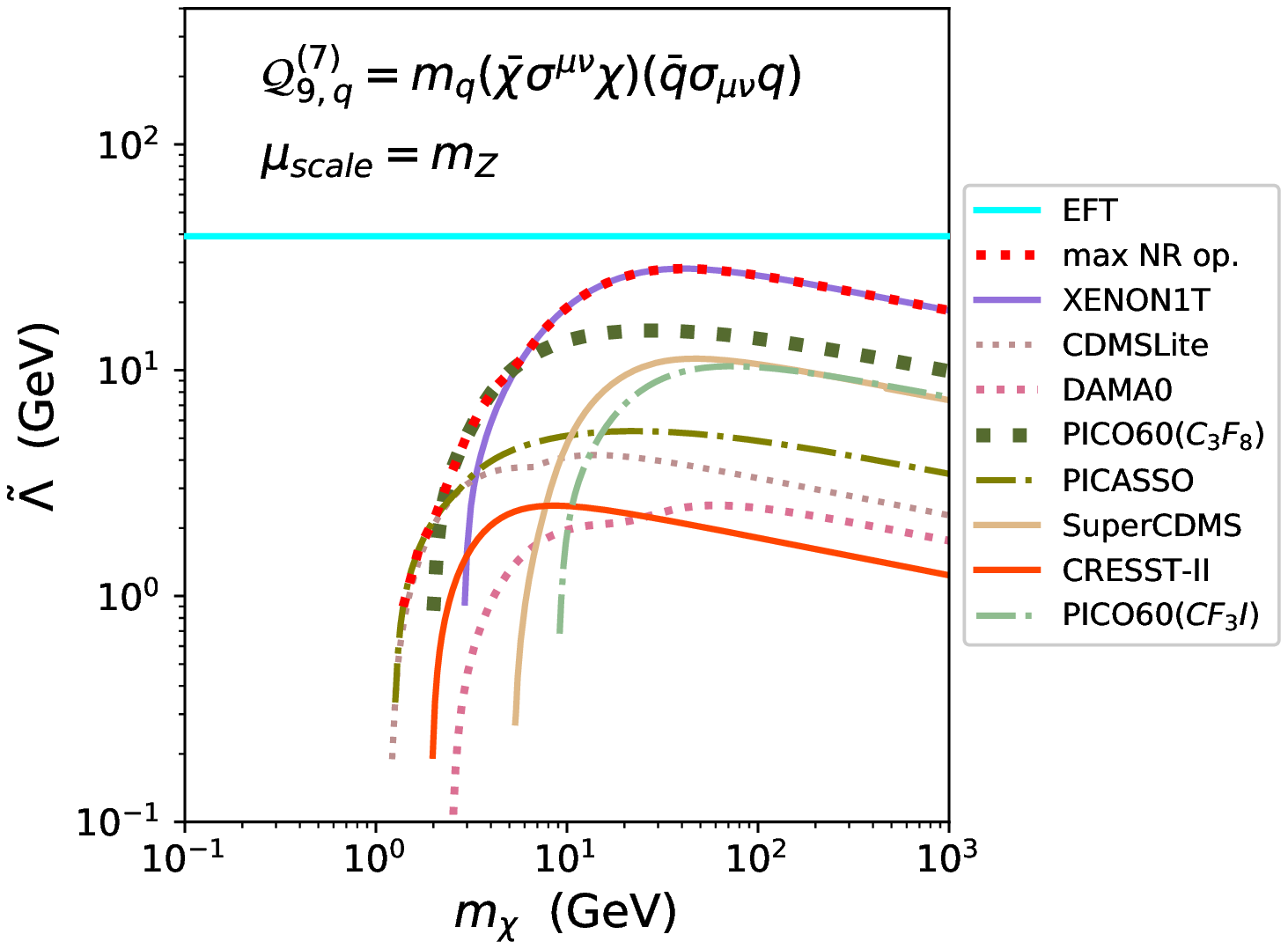}
  \includegraphics[width=0.49\columnwidth]{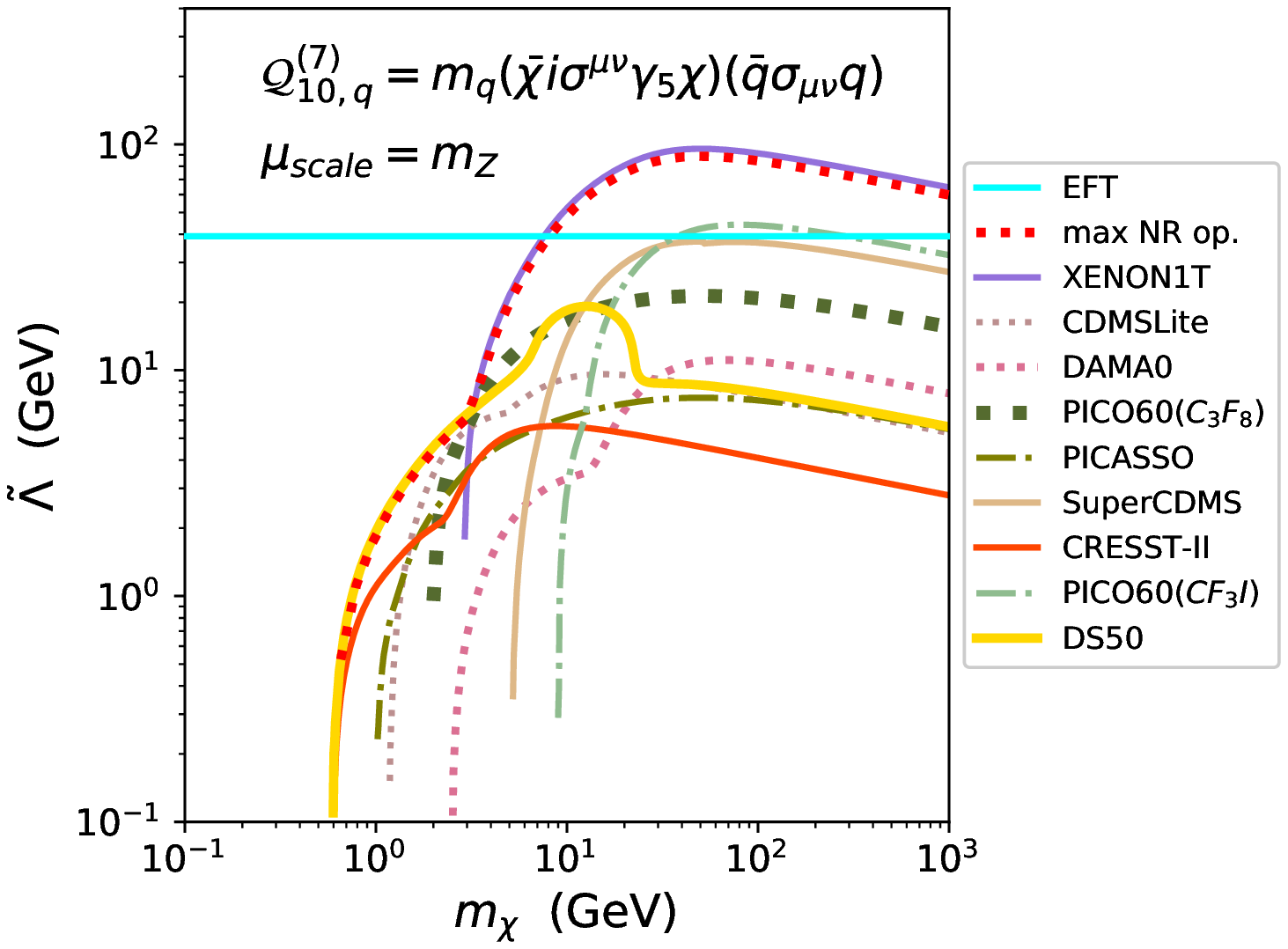}  
\end{center}
\caption{The same as in Fig.~\ref{fig:q6_12_lambda} for $\Q^{(7)}_{9}$
  {\bf (left)} and $\Q^{(7)}_{10}$ {\bf (right)}.}
\label{fig:q7_9_10_lambda}
\end{figure}

In all the plots the data are analyzed in the same way of the
experimental collaborations to obtain lower bounds on the effective
scale $\tilde{\Lambda}$ defined in Eq.~(17) as a function of the WIMP
mass $m_{\chi}$ assuming a single flavor--independent coupling
$\C_{a,q}^{(d)}$ common to all quarks. In particular, for all
experiments with the exception of SuperCDMS and DarkSide--50 we
compare the expected rate to the 90\% C.L. upper bound on the count
rate in each energy bin assuming zero background. Namely, for XENON1T
we have assumed 7 WIMP candidate events in the range of 3 PE $ \le S_1
\le $ 70 PE, as shown in Fig.~3 of Ref.~\cite{xenon_2018} for the
primary scintillation signal S1 (directly in Photo Electrons, PE),
with an exposure of 278.8 days, fiducial volume of 1.3 ton and the
efficiency taken from Fig.~1 of Ref.~\cite{xenon_2018}. In the
analysis of DarkSide--50, we subtract the estimated background by
fitting the data at fixed $m_{\chi}$ to the sum
$S_i(\tilde{\Lambda})+\lambda b_i$ in terms of the two free parameters
$\tilde{\Lambda}$ and $\lambda$~\cite{sogang_scaling_law_nr}, with
$S_i$ the expected WIMP signal in each energy bin $i$ and $b_i$ taken
from Fig.~3 of~\cite{ds50} using the exposure of 6786.0 kg days. This latter
procedure is particularly effective when the spectral shapes of the
signal and of the background are different.  For DarkSide--50 the
estimated spectrum of the background is rising with the recoil energy,
so it yields a weaker constraint for interactions types with an
explicit momentum dependence that lead to a signal rising with energy
in a way similar to the background. This loss of constraining power is
the reason of the peculiar shapes of some of the exclusion plots for
DarkSide--50 in Figs.~\ref{fig:q5_lambda}--\ref{fig:q7_9_10_lambda}.
The latest SuperCDMS analysis~\cite{super_cdms_2017} observed 1 event
between 4 and 100 keVnr with an exposure of 1690 kg days. 
To analyze the observed spectrum we apply the 
the maximum--gap method~\cite{yellin} with the efficiency taken from 
Fig.~1 of~\cite{super_cdms_2017} and the energy resolution
$\sigma=\sqrt{0.293^2+0.056^2 E_{ee}/\rm keVee}~\rm keVee$ from~\cite{cdms_resolution}.
In the case of CDMSlite, we consider the energy bin 0.056
keV$<E^{\prime}<$ 1.1 keV with a measured count rate of 1.1$\pm$0.2
[keV kg days]$^{-1}$ (Full Run 2 rate, Table II of
Ref.~\cite{cdmslite_2017}). We have taken the efficiency from Fig.~4
of~\cite{cdmslite_2017} and the energy resolution
$\sigma=\sqrt{\sigma_E^2+B E_R+(A E_R)^2}$, with $\sigma_E$=9.26 eV,
$A$=5.68$\times 10^{-3}$ and $B$=0.64 eV from Section IV.A
of~\cite{cdmslite_2017}. In the case of threshold detectors such as
PICO60 and PICASSO, we consider for each threshold an energy bin up to
the maximal recoil energy allowed by the escape velocity.  For PICASSO
we take into account the six energy thresholds ($E_{th}$=1.0, 1.5, 2.7,
6.6,15.7,36.8 keV) analyzed in~\cite{picasso} while for PICO-60 we
considered the threshold $E_{th}$=3.3 keV with a total exposure of
1167.0 kg days and no event detected~\cite{pico60}.  PICO-60 can also
employ a $CF_3I$ target and in this case we adopt an energy threshold
of 13.6 keV and an exposure of 1335 kg days~\cite{pico60_2015}.  For
DAMA we consider the upper bound from the average count rate (DAMA0)
which has been taken from~\cite{damaz} (rebinned from 0.25-keVee- to
0.5-keVee-width bins).  We assume constant quenching factors $q$=0.3
for sodium and $q$=0.09 for iodine, and the energy resolution $\sigma$
= 0.0091 (E$_{ee}$/keVee) + 0.448 $\sqrt{E_{ee}/{\rm keVee}}$ in keV.
For the CRESST-II experiment, we considered the Lise module analysis
from~\cite{cresst_II} with energy resolution $\sigma$=0.062 keV and
detector efficiency from Fig. 4 of~\cite{cresst_II_description}. In
our analysis, we have selected 15 events for 0.3 keVnr$<E_R<$ 0.49
keVnr with an exposure of 52.15 kg days. With the assumptions
summarized above we reproduce published results in the case of a
standard Spin--Independent interaction.  Further details can be found
in Appendix A of Ref.~\cite{sogang_scaling_law_nr}.


The exclusion plots of
Figs.~\ref{fig:q5_lambda}--\ref{fig:q7_9_10_lambda} can be roughly
devided in two classes: in the case of models $\Q^{(5)}_{1,q}$,
$\Q^{(5)}_{2,q}$, $\Q^{(6)}_{1,q}$, $\Q^{(6)}_{2,q}$, $\Q^{(7)}_{1}$,
$\Q^{(7)}_{2}$, $\Q^{(7)}_{5,q}$, $\Q^{(7)}_{6,q}$ and
$\Q^{(7)}_{10,q}$ the most constraining experiments are DarkSide--50
at low WIMP mass and XENON1T at larger $m_{\chi}$. As can be seen by
combining Table~\ref{table:interferences} (that allows to see the
correspondence between each $\Q^{(d)}_{a,q}$, $\Q^{(d)}_{b}$ term and
NR operators) and Table~\ref{table:eft_summary} (where the
correspondence between each NR operator ${O}_i$ and the nuclear
response functions $W^{\tau\tau^{\prime}}_{T k}(y)$'s is shown) one
can see that all such interactions take contributions from the $W_M$
nuclear response function, leading to a SI scaling of the cross
section (possibly combined with explicit dependence from the exchanged
momentum $q^2$ and from the WIMP incoming speed~\footnote{As far as
  the ${\cal O}_5$ and ${\cal O}_8$ NR operators are concerned, the SI
  part of the nuclear response function usually dominates in spite of
  the fact that it is velocity suppressed
  \cite{sogang_scaling_law_nr}.}). Indeed, due to its very low
threshold DarkSide--50 drives the exclusion plot at low mass, but only
for interactions that do not require a nuclear spin (its target is
$^{40}$Ar), while at larger masses the SI coupling enhances the
sensitivity for scatterings off xenon in XENON1T. The second class of
exclusion plots is represented by the models $\Q^{(6)}_{3,q}$,
$\Q^{(6)}_{4,q}$, $\Q^{(7)}_{3}$, $\Q^{(7)}_{4}$, $\Q^{(7)}_{7,q}$,
$\Q^{(7)}_{8,q}$, and $\Q^{(7)}_{9,q}$ for which the exclusion plot is
driven by PICASSO and PICO60 (and, sometimes, by CDMSLite) at low WIMP
mass, and by XENON1T at larger WIMP masses. In such cases, as can
again be seen from Tables~\ref{table:interferences}
and~\ref{table:eft_summary}, the operator $M$ is always missing in the
NR limit, while the response functions $\Sigma^{\prime}$ and/or
$\Sigma^{\prime\prime}$ are always present, leading to a SD--type
scaling of the cross section for which large detectors containing
fluorine are competitive with xenon.

Some of the limits shown in
Figs.~\ref{fig:q5_lambda}--\ref{fig:q7_9_10_lambda} may be so weak
that they put bounds on values of the $\tilde{\Lambda}$ scale which
are inconsistent with the validity of the effective theory. In such
case one can simply conclude that the present experimental sensitivity
of direct detection experiments is not able to put bounds on the
corresponding effective operator. A criterion for the validity of the
EFT is to interpret the scale $\tilde{\Lambda}$ in terms of a
propagator $g^2/M_{*}^2$ with $g<\sqrt{4\pi}$ and $M_{*}>\mu_{scale}$,
since in our analysis we fixed the boundary conditions of the EFT at
the scale $\mu_{scale}$. This is straightforward for dimension--6
operators, while in the case of operators whose effective coupling has
dimension different from -2 only matching the EFT with the full theory
would allow to draw robust conclusions. In particular, in this case
$\tilde{\Lambda}$ can be interpreted in terms of the same propagator
times the appropriate power of a typical scale of the problem
$\mu_{scale}^{\prime}$, which depends on the ultraviolet completion of
the EFT. For instance, in the operator ${\cal Q}_{5,q}^{(7)}$=$m_q
(\bar \chi \chi)( \bar q q)$ the quark mass may originate from a
Yukawa coupling, so the missing scale is an Electroweak vacuum
expectation value in the denominator.  To fix an order of magnitude we
choose to fix $\mu_{scale}^{\prime}$ = $\mu_{scale}$, so that the
bound $\tilde{\Lambda}>\mu_{scale}/(4\pi)^{1/(d-4)}$ can be
derived. Such limit is shown as a horizontal solid line in
Figs.~\ref{fig:q5_lambda}--\ref{fig:q7_9_10_lambda}. In particular,
for models $\Q^{(7)}_{2}$, $\Q^{(7)}_{3}$, $\Q^{(7)}_{4}$,
$\Q^{(7)}_{6}$, $\Q^{(7)}_{7}$, $\Q^{(7)}_{8}$ and $\Q^{(7)}_{9}$ the
bound on the $\tilde{\Lambda}$ scale lies above such curve in all the
WIMP mass range, implying that the sensitivities of present direct
detection experiments to such couplings may not be sufficient to put
meaningful bounds. However we stress again that this can only be
assessed when a specific ultraviolet completion of the effective
theory is assumed.

In~\ref{app:program} we introduce {\verb NRDD_constraints }, a simple
interpolating code written in Python that can reproduce most of the
results of this Section by assuming that one NR coupling dominates in
the low--energy limit of the interactions of
Eqs.(\ref{eq:dim5})--(\ref{eq:dim7}). In
Figs.~\ref{fig:q5_lambda}--\ref{fig:q6_34_lambda} and
Figs.~\ref{fig:q7_12_lambda}--\ref{fig:q7_9_10_lambda} the output of
such code is indicated by ``max NR op.'' and represented by the
red--dashed curve. One can see that, with the exception of models
${\cal Q}_{7,q}^{(7)}$ and ${\cal Q}_{8,q}^{(7)}$, the ``max NR op.''
matches the lower edge of the excluded region obtained through a full
calculation.

\section{Interference and momentum effects in the NR theory}
\label{sec:interferences}

No matter which among the relativistic interactions listed in
Eqs.(\ref{eq:dim5},\ref{eq:dim6},\ref{eq:dim7}) is generated at a
higher scale by some beyond--the--standard--model scenario, Dark
Matter DD scattering is a low--energy process completely described by
the NR effective theory of (Eq.~\ref{eq:H}). This implies that the
limits discussed in the previous Section can be expressed in terms of
NR operators only. In particular, in the case of interactions between
the DM particle and the quark current, this would have the advantage
to present the limits from existing experiments in a way independent
from the choice of the $\C_{a,q}^{(d)}$ couplings for each flavor $q$,
since the NR effective theory depends only on WIMP mass and on the
ratio between the WIMP--neutron and the WIMP--proton couplings
$r\equiv c^n/c^p$. Indeed, in ref. \cite{sogang_scaling_law_nr} we
obtained updated upper bounds on the effective cross section:

\begin{equation}
\sigma^{\cal N}_{i}=\max(\sigma^{p}_{i},\sigma^{n}_{i}),
  \label{eq:conventional_sigma_nucleon_old}
\end{equation}

\noindent with:

\begin{equation}
\sigma^{p,n}_{i}=(c_{i}^{p,n})^2\frac{\mu_{\chi{\cal N}}^2}{\pi},
  \label{eq:conventional_sigma_old}
\end{equation}

\noindent ($\mu_{\chi{\cal N}}$ is the WIMP--nucleon reduced mass)
assuming constant couplings $c_{i}^{p,n}$ and, systematically, dominance
of one of the possible NR interaction terms $\op_i$ of
Eq. (\ref{eq:H}), providing for each of them a two--dimensional plot
where the contours of the most stringent 90\% C.L. upper bounds to
$\sigma^{\cal N}_{i}$ were shown as a function of the two parameters
$m_{\chi}$ (WIMP mass), and $c^n/c^p$.  One possible drawback of this
approach is however that, in general, a given relativistic coupling
leads to more than one NR operator.  In addition to that, as explained
in Section \ref{sec:rel_eft}, the NR coefficients $c_i^{\tau}$ may
depend explicitly on the exchanged momentum, leading, in practice, to
contributions which are equivalent to including additional NR
operators of the type $F_i^{\alpha}(q^2){\cal O}_i$ (where, for each
operator ${\cal O}_i$ different momentum dependences are possible, as
for instance in Eqs.(\ref{eq:F_PP'},\ref{eq:F_tildeG})). In fact,
setting:

\begin{eqnarray}
  c_{i}^{\tau}(m_{\chi},q^2)&\equiv&
  \hat{c}_{i,\alpha}^{\tau}(m_{\chi})F_i^{\alpha}(q^2)\,,\\ R_k^{\tau\tau^{\prime}}&\equiv&
  c^{\tau}_i c^{\tau'}_j
  \hat{R}^{\tau\tau^{\prime}}_{k,ij}=\hat{c}^{\tau}_{i,\alpha}
  \hat{c}^{\tau'}_{j,\beta}
  \hat{R}^{\tau\tau^{\prime}}_{k,ij}F_i^{\alpha}(q^2)F_j^{\beta}(q^2)\,,
  \label{eq:cr_factorizations}
\end{eqnarray}

\noindent the squared amplitude
(\ref{eq:squared_amplitude}) can be rewritten as:
\begin{eqnarray}
  \frac{1}{2 j_{\chi}+1} \frac{1}{2 j_{T}+1}|\mathcal{M}_T|^2&=&
  \frac{4\pi}{2 j_{T}+1} \sum_{ij} \sum_{\alpha\beta}\sum_{\tau,\tau'} \hat{c}_{i,\alpha}^{\tau}\hat{c}_{j,\beta}^{\tau^{\prime}}\left [\sum_k \hat{R}^{\tau\tau^{\prime}}_{k,ij} W^{\tau\tau^{\prime}}_{k}(q^2)\right ]F_i^{\alpha}(q^2)F_j^{\beta}(q^2),\nonumber
  \end{eqnarray}
\noindent so that the expected rate $R$ can be expressed as a sum over
all possible interferences among the contributions from each
generalized NR term $F_i^{\alpha}(q^2){\cal O}_i$:

\begin{equation}
R=\sum_{ij} \sum_{\alpha\beta}\sum_{\tau,\tau'}\hat{c}_{i,\alpha}^{\tau}\hat{c}_{j,\beta}^{\tau^{\prime}}\langle{\cal
  O}_i{\cal O}_jF_i^{\alpha}(q^2)F_j^{\beta}(q^2) \rangle_{\tau\tau^{\prime}}.
\label{eq:interferences}
  \end{equation}

In the equation above each term $\langle{\cal O}_i{\cal
  O}_jF_i^{\alpha}(q^2)F_j^{\beta}(q^2) \rangle_{\tau\tau^{\prime}}$
simply represents the factor that multiplies
$\hat{c}_{i,\alpha}^{\tau}\hat{c}_{j,\beta}^{\tau^{\prime}}$ at fixed
$i$, $j$, $\alpha$, $\beta$, $\tau$, $\tau^{\prime}$ in the expected
rate. The terms contributing to the sums over $i$, $j$, $\alpha$,
$\beta$ for each of the interactions discussed in
Section~\ref{sec:analysis} are listed in
Table~\ref{table:interferences}.

\begin{table}[t]
\begin{center}
  {\begin{tabular}{@{}|c|c|@{}}
\hline      
$\Q_{1}^{(5)}$  & $\hat{c}_{1}^{\tau}\hat{c}_{1}^{\tau^{\prime}}\langle{\cal
  O}_1{\cal O}_1\rangle_{\tau\tau^{\prime}}+\hat{c}_{5}^{\tau}\hat{c}_{5}^{\tau^{\prime}}\langle{\cal
  O}_5{\cal O}_5\frac{1}{q^4}\rangle_{\tau\tau^{\prime}}+\hat{c}_{4}^{\tau}\hat{c}_{4}^{\tau^{\prime}}\langle{\cal
  O}_4{\cal O}_4\rangle_{\tau\tau^{\prime}}$\\
   & $+\hat{c}_{6}^{\tau}\hat{c}_{6}^{\tau^{\prime}}\langle{\cal
  O}_6{\cal O}_6 \frac{1}{q^4}\rangle_{\tau\tau^{\prime}}+\hat{c}_{4}^{\tau}\hat{c}_{5}^{\tau^{\prime}}\langle{\cal
  O}_4{\cal O}_5 \frac{1}{q^2}\rangle_{\tau\tau^{\prime}}+\hat{c}_{4}^{\tau}\hat{c}_{6}^{\tau^{\prime}}\langle{\cal
  O}_4{\cal O}_6 \frac{1}{q^2}\rangle_{\tau\tau^{\prime}}$\\
\hline      
$\Q_2^{(5)}$  & $\hat{c}_{11}^{\tau}\hat{c}_{11}^{\tau^{\prime}}\langle{\cal
  O}_{11}{\cal O}_{11}\frac{1}{q^4} \rangle_{\tau\tau^{\prime}}$ \\
\hline      
$\Q_1^{(6)}$  & $\hat{c}_{1}^{\tau}\hat{c}_{1}^{\tau^{\prime}}\langle{\cal
  O}_1{\cal O}_1\rangle_{\tau\tau^{\prime}}$ \\
\hline      
$\Q_2^{(6)}$  & $\hat{c}_{8}^{\tau}\hat{c}_{8}^{\tau^{\prime}}\langle{\cal
  O}_8{\cal O}_8\rangle_{\tau\tau^{\prime}}+\hat{c}_{9}^{\tau}\hat{c}_{9}^{\tau^{\prime}}\langle{\cal
  O}_9{\cal O}_9\rangle_{\tau\tau^{\prime}}+\hat{c}_{8}^{\tau}\hat{c}_{9}^{\tau^{\prime}}\langle{\cal
  O}_8{\cal O}_9\rangle_{\tau\tau^{\prime}}$\\
\hline      
$\Q_3^{(6)}$  & $\hat{c}_{7}^{\tau}\hat{c}_{7}^{\tau^{\prime}}\langle{\cal
  O}_7{\cal O}_7\rangle_{\tau\tau^{\prime}}+\hat{c}_{9}^{\tau}\hat{c}_{9}^{\tau^{\prime}}\langle{\cal
  O}_9{\cal O}_9\rangle_{\tau\tau^{\prime}}$ \\
\hline      
$\Q_4^{(6)}$  & $\hat{c}_{4}^{\tau}\hat{c}_{4}^{\tau^{\prime}}\langle{\cal
  O}_4{\cal O}_4\rangle_{\tau\tau^{\prime}}+\hat{c}_{6}^{\tau}\hat{c}_{6}^{\tau^{\prime}}\langle{\cal
  O}_6{\cal O}_6\rangle_{\tau\tau^{\prime}}$ \\
  &$+\hat{c}_{6}^{\tau}\hat{c}_{6}^{\tau^{\prime}}\langle{\cal
  O}_6{\cal O}_6\frac{m^4_N}{(m^2_\pi-q^2)^2}\rangle_{\tau\tau^{\prime}}+\hat{c}_{6}^{\tau}\hat{c}_{6}^{\tau^{\prime}}\langle{\cal
  O}_6{\cal O}_6\frac{m^4_N}{(m^2_\pi-q^2)(m^2_\eta-q^2)}\rangle_{\tau\tau^{\prime}}$\\
  &$+\hat{c}_{6}^{\tau}\hat{c}_{6}^{\tau^{\prime}}\langle{\cal
  O}_6{\cal O}_6\frac{m^4_N}{(m^2_\eta-q^2)^2}\rangle_{\tau\tau^{\prime}}+\hat{c}_{6}^{\tau}\hat{c}_{6}^{\tau^{\prime}}\langle{\cal
  O}_6{\cal O}_6\frac{m^2_N}{(m^2_\pi-q^2)}\rangle_{\tau\tau^{\prime}}+\hat{c}_{6}^{\tau}\hat{c}_{6}^{\tau^{\prime}}\langle{\cal
  O}_6{\cal O}_6\frac{m^2_N}{(m^2_\eta-q^2)}\rangle_{\tau\tau^{\prime}}$\\
  &$+\hat{c}_{4}^{\tau}\hat{c}_{6}^{\tau^{\prime}}\langle{\cal
  O}_4{\cal O}_6\frac{m^2_N}{(m^2_\pi-q^2)}\rangle_{\tau\tau^{\prime}}+\hat{c}_{4}^{\tau}\hat{c}_{6}^{\tau^{\prime}}\langle{\cal
  O}_4{\cal O}_6\frac{m^2_N}{(m^2_\eta-q^2)}\rangle_{\tau\tau^{\prime}}+\hat{c}_{4}^{\tau}\hat{c}_{6}^{\tau^{\prime}}\langle{\cal
  O}_4{\cal O}_6\rangle_{\tau\tau^{\prime}}$ \\
\hline      
$\Q_1^{(7)}$  & $\hat{c}_{1}^{\tau}\hat{c}_{1}^{\tau^{\prime}}\langle{\cal
  O}_1{\cal O}_1\rangle_{\tau\tau^{\prime}}$ \\
\hline      
$\Q_2^{(7)}$  & $\hat{c}_{11}^{\tau}\hat{c}_{11}^{\tau^{\prime}}\langle{\cal
  O}_{11}{\cal O}_{11}\rangle_{\tau\tau^{\prime}}$  \\
\hline      
$\Q_3^{(7)}$  & $\hat{c}_{10}^{\tau}\hat{c}_{10}^{\tau^{\prime}}\langle{\cal
  O}_{10}{\cal O}_{10}\rangle_{\tau\tau^{\prime}}+\hat{c}_{10}^{\tau}\hat{c}_{10}^{\tau^{\prime}}\langle{\cal
  O}_{10}{\cal O}_{10}\frac{q^4}{(m^2_\pi-q^2)^2}\rangle_{\tau\tau^{\prime}}+\hat{c}_{10}^{\tau}\hat{c}_{10}^{\tau^{\prime}}\langle{\cal
  O}_{10}{\cal O}_{10}\frac{q^4}{(m^2_\pi-q^2)(m^2_\eta-q^2)}\rangle_{\tau\tau^{\prime}}$\\
  &$+\hat{c}_{10}^{\tau}\hat{c}_{10}^{\tau^{\prime}}\langle{\cal
  O}_{10}{\cal O}_{10}\frac{q^4}{(m^2_\eta-q^2)^2}\rangle_{\tau\tau^{\prime}}+\hat{c}_{10}^{\tau}\hat{c}_{10}^{\tau^{\prime}}\langle{\cal
  O}_{10}{\cal O}_{10}\frac{q^2}{(m^2_\pi-q^2)}\rangle_{\tau\tau^{\prime}}+\hat{c}_{10}^{\tau}\hat{c}_{10}^{\tau^{\prime}}\langle{\cal
  O}_{10}{\cal O}_{10}\frac{q^2}{(m^2_\eta-q^2)}\rangle_{\tau\tau^{\prime}}$ \\
\hline      
$\Q_4^{(7)}$  & $\hat{c}_{6}^{\tau}\hat{c}_{6}^{\tau^{\prime}}\langle{\cal
  O}_{6}{\cal O}_{6}\rangle_{\tau\tau^{\prime}}+\hat{c}_{6}^{\tau}\hat{c}_{6}^{\tau^{\prime}}\langle{\cal
  O}_{6}{\cal O}_{6}\frac{q^4}{(m^2_\pi-q^2)^2}\rangle_{\tau\tau^{\prime}}+\hat{c}_{6}^{\tau}\hat{c}_{6}^{\tau^{\prime}}\langle{\cal
  O}_{6}{\cal O}_{6}\frac{q^4}{(m^2_\pi-q^2)(m^2_\eta-q^2)}\rangle_{\tau\tau^{\prime}}$\\
  &$+\hat{c}_{6}^{\tau}\hat{c}_{6}^{\tau^{\prime}}\langle{\cal
  O}_{6}{\cal O}_{6}\frac{q^4}{(m^2_\eta-q^2)^2}\rangle_{\tau\tau^{\prime}}+\hat{c}_{6}^{\tau}\hat{c}_{6}^{\tau^{\prime}}\langle{\cal
  O}_{6}{\cal O}_{6}\frac{q^2}{(m^2_\pi-q^2)}\rangle_{\tau\tau^{\prime}}+\hat{c}_{6}^{\tau}\hat{c}_{6}^{\tau^{\prime}}\langle{\cal
  O}_{6}{\cal O}_{6}\frac{q^2}{(m^2_\eta-q^2)}\rangle_{\tau\tau^{\prime}}$ \\
\hline      
$\Q_5^{(7)}$  & $\hat{c}_{1}^{\tau}\hat{c}_{1}^{\tau^{\prime}}\langle{\cal
  O}_1{\cal O}_1\rangle_{\tau\tau^{\prime}}$ \\
\hline      
$\Q_6^{(7)}$  & $\hat{c}_{11}^{\tau}\hat{c}_{11}^{\tau^{\prime}}\langle{\cal
  O}_{11}{\cal O}_{11}\rangle_{\tau\tau^{\prime}}$  \\
\hline      
$\Q_7^{(7)}$  &  $\hat{c}_{10}^{\tau}\hat{c}_{10}^{\tau^{\prime}}\langle{\cal
  O}_{10}{\cal O}_{10}\rangle_{\tau\tau^{\prime}}+\hat{c}_{10}^{\tau}\hat{c}_{10}^{\tau^{\prime}}\langle{\cal
  O}_{10}{\cal O}_{10}\frac{m_N^4}{(m^2_\pi-q^2)^2}\rangle_{\tau\tau^{\prime}}+\hat{c}_{10}^{\tau}\hat{c}_{10}^{\tau^{\prime}}\langle{\cal
  O}_{10}{\cal O}_{10}\frac{m_N^4}{(m^2_\pi-q^2)(m^2_\eta-q^2)}\rangle_{\tau\tau^{\prime}}$\\
  &$+\hat{c}_{10}^{\tau}\hat{c}_{10}^{\tau^{\prime}}\langle{\cal
  O}_{10}{\cal O}_{10}\frac{m_N^4}{(m^2_\eta-q^2)^2}\rangle_{\tau\tau^{\prime}}+\hat{c}_{10}^{\tau}\hat{c}_{10}^{\tau^{\prime}}\langle{\cal
  O}_{10}{\cal O}_{10}\frac{m_N^2}{(m^2_\pi-q^2)}\rangle_{\tau\tau^{\prime}}+\hat{c}_{10}^{\tau}\hat{c}_{10}^{\tau^{\prime}}\langle{\cal
  O}_{10}{\cal O}_{10}\frac{m_N^2}{(m^2_\eta-q^2)}\rangle_{\tau\tau^{\prime}}$\\
\hline      
$\Q_8^{(7)}$  &  $\hat{c}_{6}^{\tau}\hat{c}_{6}^{\tau^{\prime}}\langle{\cal
  O}_{6}{\cal O}_{6}\rangle_{\tau\tau^{\prime}}+\hat{c}_{6}^{\tau}\hat{c}_{6}^{\tau^{\prime}}\langle{\cal
  O}_{6}{\cal O}_{6}\frac{m_N^4}{(m^2_\pi-q^2)^2}\rangle_{\tau\tau^{\prime}}+\hat{c}_{6}^{\tau}\hat{c}_{6}^{\tau^{\prime}}\langle{\cal
  O}_{6}{\cal O}_{6}\frac{m_N^4}{(m^2_\pi-q^2)(m^2_\eta-q^2)}\rangle_{\tau\tau^{\prime}}$\\
  &$+\hat{c}_{6}^{\tau}\hat{c}_{6}^{\tau^{\prime}}\langle{\cal
  O}_{6}{\cal O}_{6}\frac{m_N^4}{(m^2_\eta-q^2)^2}\rangle_{\tau\tau^{\prime}}+\hat{c}_{6}^{\tau}\hat{c}_{6}^{\tau^{\prime}}\langle{\cal
  O}_{6}{\cal O}_{6}\frac{m_N^2}{(m^2_\pi-q^2)}\rangle_{\tau\tau^{\prime}}+\hat{c}_{6}^{\tau}\hat{c}_{6}^{\tau^{\prime}}\langle{\cal
  O}_{6}{\cal O}_{6}\frac{m_N^2}{(m^2_\eta-q^2)}\rangle_{\tau\tau^{\prime}}$ \\
\hline      
$\Q_9^{(7)}$  & $\hat{c}_{4}^{\tau}\hat{c}_{4}^{\tau^{\prime}}\langle{\cal
  O}_4{\cal O}_4\rangle_{\tau\tau^{\prime}}$ \\
\hline      
$\Q_{10}^{(7)}$  & $\hat{c}_{10}^{\tau}\hat{c}_{10}^{\tau^{\prime}}\langle{\cal
  O}_{10}{\cal O}_{10}\rangle_{\tau\tau^{\prime}} +\hat{c}_{11}^{\tau}\hat{c}_{11}^{\tau^{\prime}}\langle{\cal
  O}_{11}{\cal O}_{11}\rangle_{\tau\tau^{\prime}} +\hat{c}_{12}^{\tau}\hat{c}_{12}^{\tau^{\prime}}\langle{\cal
  O}_{12}{\cal O}_{12}\rangle_{\tau\tau^{\prime}}$ \\
  & $+\hat{c}_{11}^{\tau}\hat{c}_{12}^{\tau^{\prime}}\langle{\cal
  O}_{11}{\cal O}_{12}\rangle_{\tau\tau^{\prime}}$\\
\hline
\end{tabular}}
\caption{Non--relativistic interaction terms contributing to the
  direct detection expected rate for each of the interactions listed
  in Eqs.(\ref{eq:dim5}--\ref{eq:dim7}).
  \label{table:interferences}}
\end{center}
\end{table}


\begin{figure}
\begin{center}
  \includegraphics[width=0.45\columnwidth]{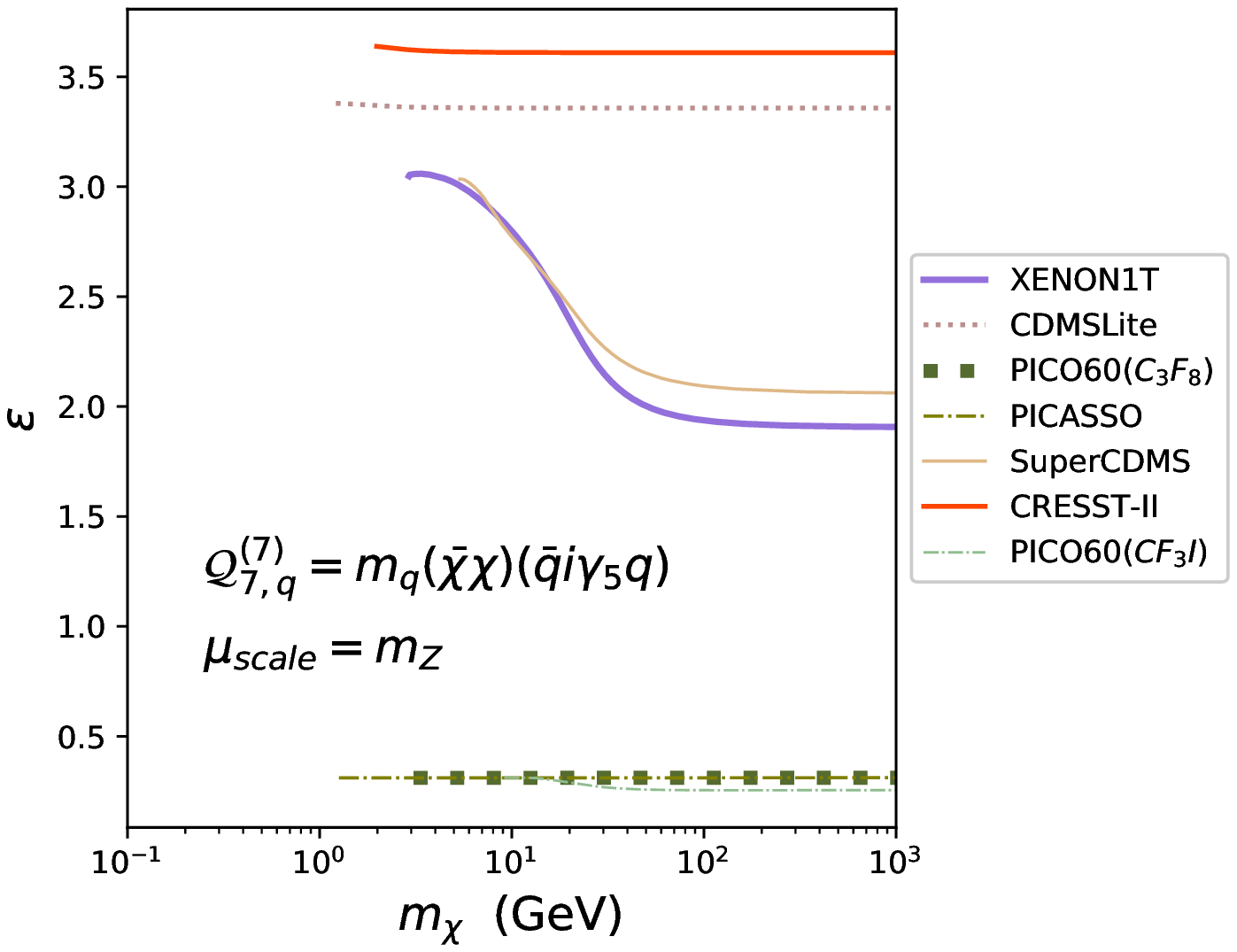}
  \includegraphics[width=0.54\columnwidth]{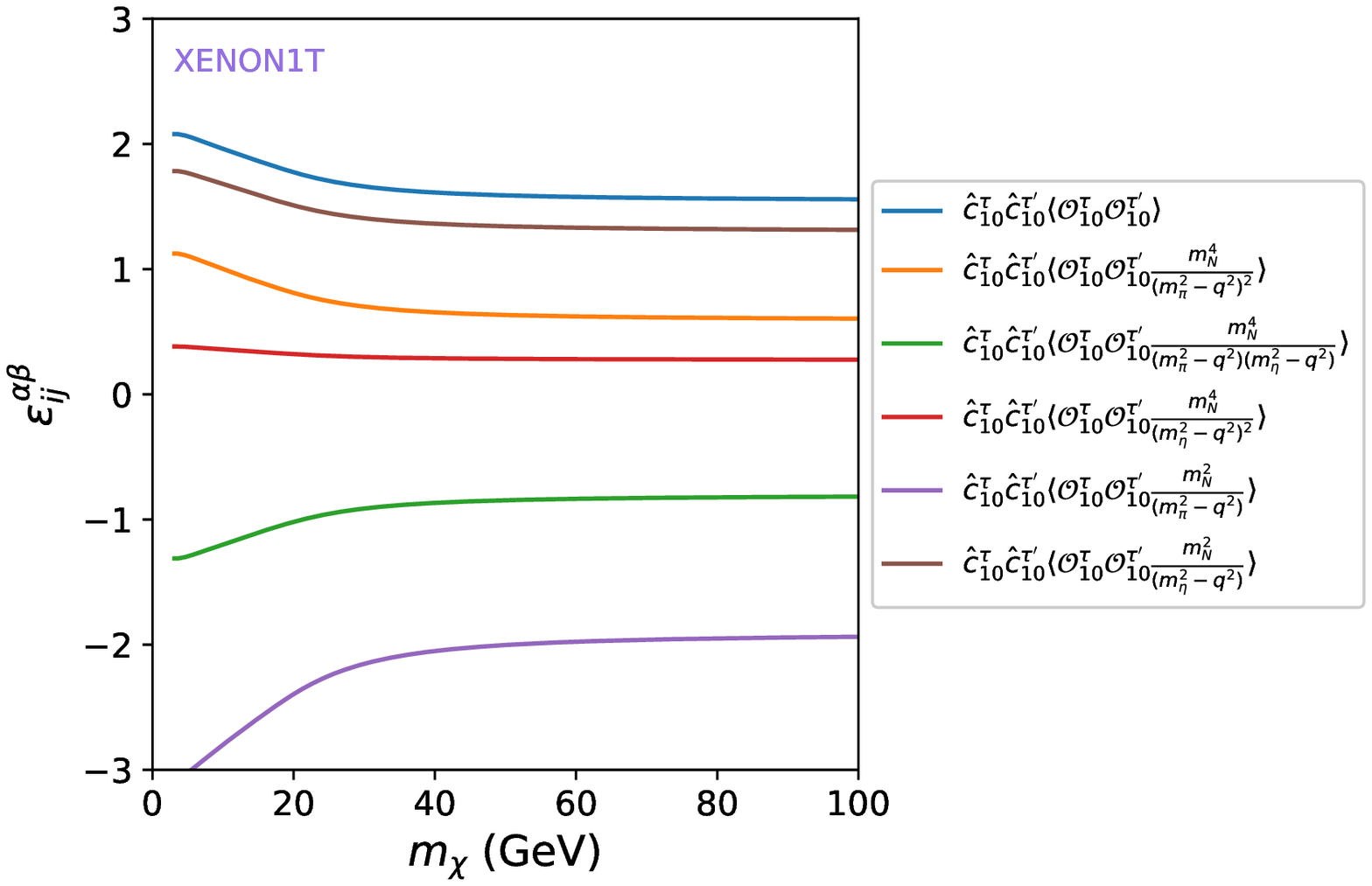}
\end{center}
\caption{Parameter $\epsilon$ defined in Eq.~(\ref{eq:epsilon}) as a
  function of the WIMP mass $m_{\chi}$ for model
  $\Q^{(7)}_{7,q}$. {\bf (left)} Parameter $\epsilon$ as a function of the
  WIMP mass $m_{\chi}$ for all the experiments and energy bins
  considered in Section \ref{sec:analysis}. {\bf (right)}
  Contributions $\epsilon_{ij}^{\alpha\beta}$ (each arising from one
  of the terms listed in Table~\ref{table:interferences}) for the
  specific example of the XENON1T experiment.}
\label{fig:epsilon_q7_7}
\end{figure}

\begin{figure}
\begin{center}
  \includegraphics[width=0.45\columnwidth]{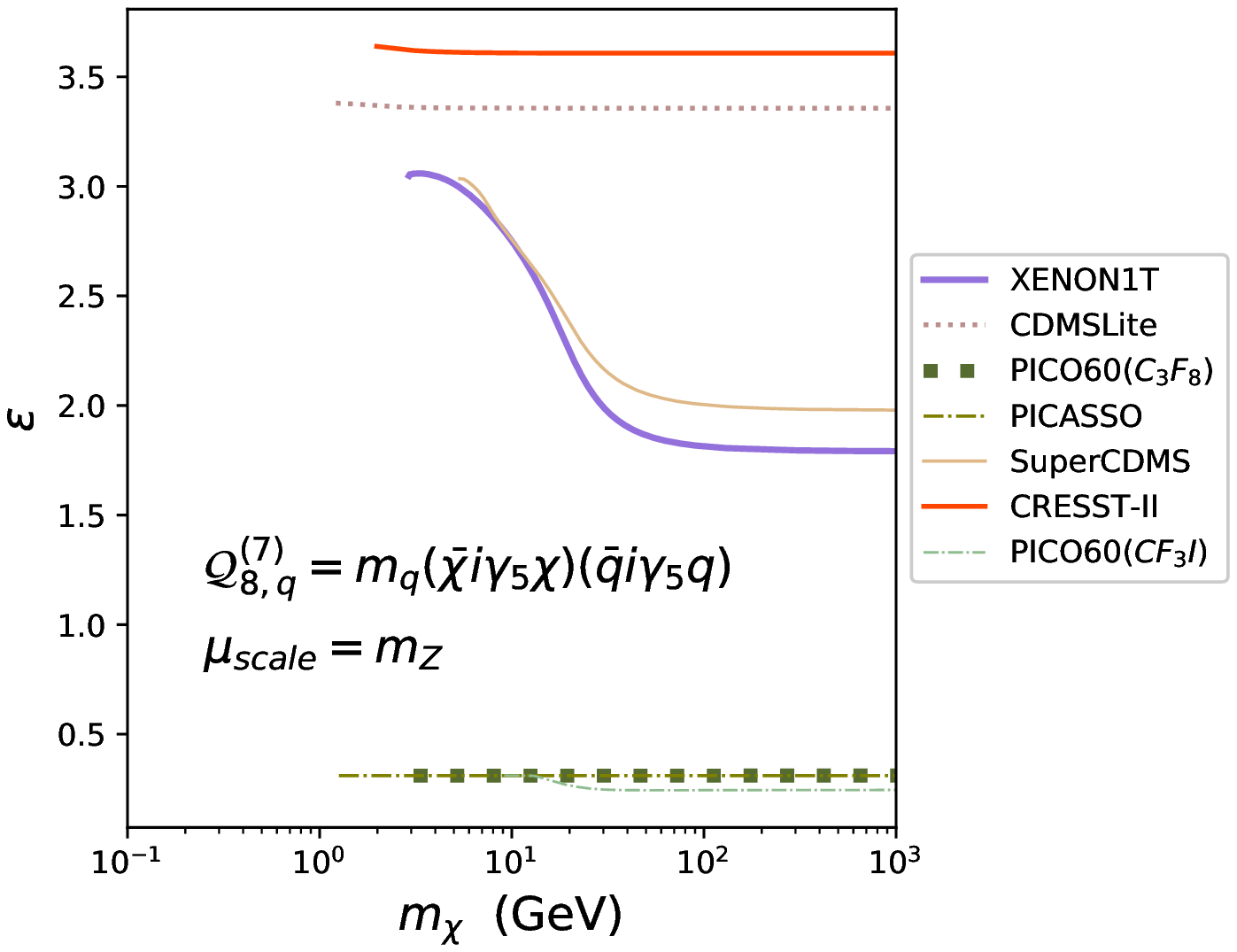}
  \includegraphics[width=0.525\columnwidth]{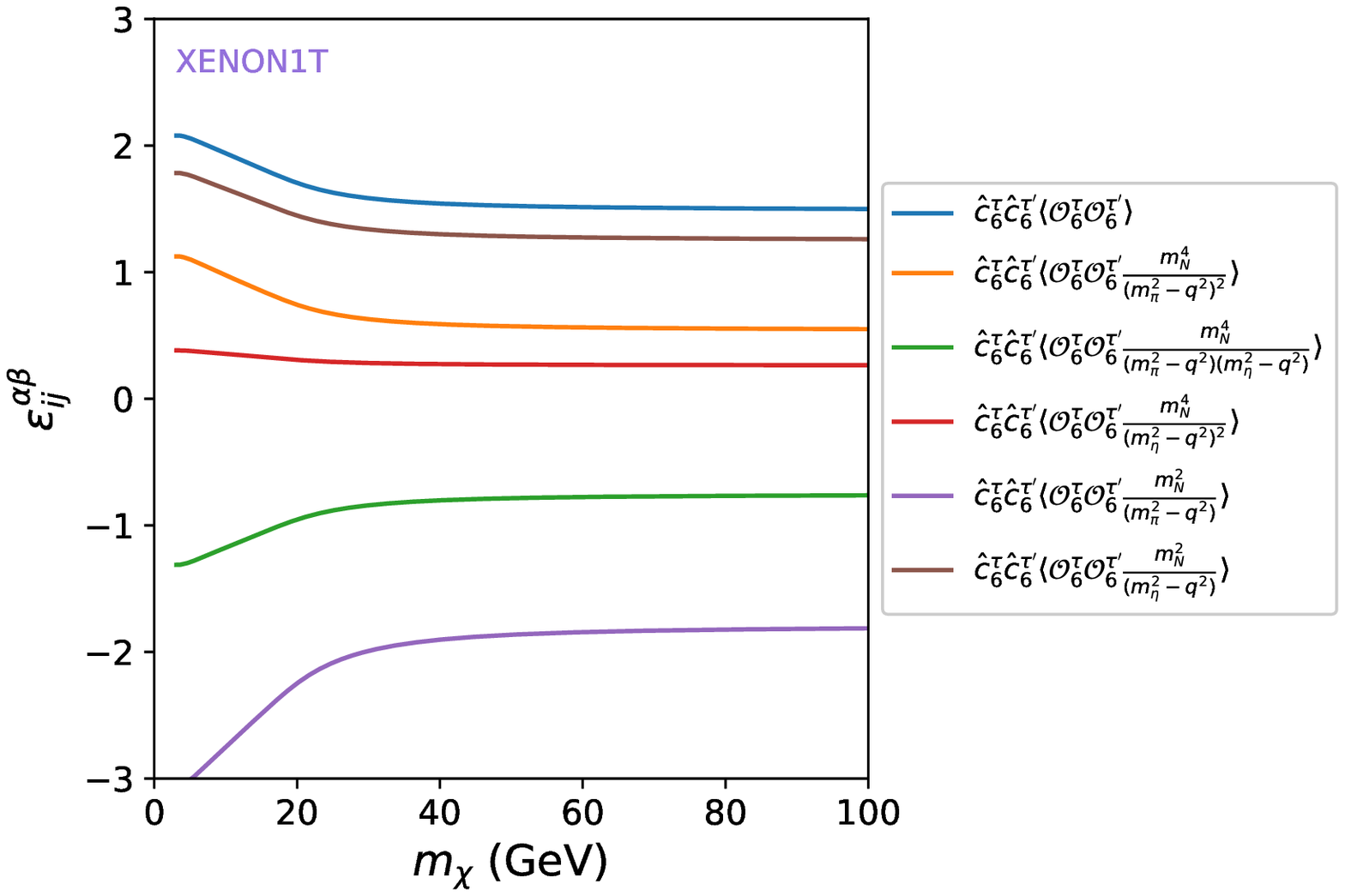}
\end{center}
\caption{The same as in Fig.~\ref{fig:epsilon_q7_7} for model
  $\Q^{(7)}_{8,q}$.}
\label{fig:epsilon_q7_8}
\end{figure}
To discuss whether it is correct to assume dominance of one effective
operator $F_i^{\alpha}(q^2){\cal O}_i$ at a time we introduce the
parameters:

\begin{equation}
\epsilon_{ij}^{\alpha\beta}=\frac{\sum_{\tau\tau^{\prime}}\hat{c}_{i,\alpha}^{\tau}\hat{c}_{j,\beta}^{\tau^{\prime}}\langle{\cal O}_i{\cal
    O}_jF_i^{\alpha}(q^2)F_j^{\beta}(q^2) \rangle_{\tau\tau^{\prime}}}{\sum_{\tau\tau^{\prime}}\sum_{lm}\sum_{\rho\sigma}\hat{c}_{l,\rho}^{\tau}\hat{c}_{m,\sigma}^{\tau^{\prime}}\langle{\cal O}_l{\cal
    O}_mF_l^{\rho}(q^2)F_m^{\sigma}(q^2) \rangle_{\tau\tau^{\prime}}},\,\,\,\, \epsilon\equiv\max_{i,j,\alpha,\beta}(|\epsilon_{ij}^{\alpha\beta}|).
  \label{eq:epsilon}
\end{equation}

\noindent
By numerical inspection we find that, with the exception of the
operators $\Q^{(7)}_{7,q}$ and $\Q^{(7)}_{8,q}$, the $\epsilon$
parameter in all the energy bins of all the experiments included in
our analysis never exceeds 1.7. Actually, we have checked that such
extreme value (indicating destructive interference) corresponds to the
highest energy bins of DAMA0 where the rate is exponentially
suppressed by the velocity distribution and irrelevant for the
constraint. In all other cases $\epsilon\lsim$1.3.  As a consequence,
assuming dominance of one of the combinations $F_i^{\alpha}(q^2){\cal
  O}_i$ in the calculation of the expected rate in the determination
of the exclusion plot implies an inaccuracy within $\pm \simeq$ 40-50
\% for $\epsilon>1$ and a factor of 2 for $\epsilon<1$, but in most
cases much smaller.

In Figs. \ref{fig:epsilon_q7_7} and \ref{fig:epsilon_q7_8} we plot
$\epsilon$ as a function of the WIMP mass $m_{\chi}$ for the two
operators $\Q^{(7)}_{7,q}$ and $\Q^{(7)}_{8,q}$.  In particular, while
the dominant contribution for $\Q^{(7)}_3$ and $\Q^{(7)}_4$
corresponds to the constant term in Eq.(\ref{eq:F_tildeG}), as shown
in Figs. \ref{fig:epsilon_q7_7} and \ref{fig:epsilon_q7_8} the
situation is different for $\Q^{(7)}_7$ and $\Q^{(7)}_8$, where each
of the terms ${\cal O}_n$,${\cal
  O}_n\frac{m_N^2}{m_{\pi}^2-q^2}$,${\cal
  O}_n\frac{m_N^2}{m_{\eta}^2-q^2}$ (with $n$=10,6) is of the same
size with large cancellations among them (as indicated by
$\epsilon\gg$ 1 values in the left--hand plot). This is confirmed by
the right--hand plot of each of the two figures, where we show
explicitly the $\epsilon_{ij}^{\alpha\beta}$ contributions for the
specific example of the XENON1T experiment. Indeed for the interaction
terms $\Q^{(7)}_7$ and $\Q^{(7)}_8$ such effect is natural since in
this particular case the momentum--independent term ${\cal O}_n$ is
next-to-leading order in chiral counting~\cite{bishara_2017} compared
to the terms ${\cal O}_n\frac{m_N^2}{m_{\pi}^2-q^2}$ and ${\cal
  O}_n\frac{m_N^2}{m_{\eta}^2-q^2}$.

Indeed, our conclusions on the $\epsilon$ parameter are not
unexpected, since the scaling of the rate for different NR operators
is very different, and depends also on experimental inputs, so barring
accidental cancellations or clear--cut situations, like the one of
axial operators $\Q^{(7)}_{7,q}$ and $\Q^{(7)}_{8,q}$,
dominance of one NR operator appears natural. The numerical tests in
this Section confirm this. In~\ref{app:program} we introduce
{\verb NRDD_constraints }, a simple code that exploits this feature to
calculate approximate bounds on the couplings $\C^{(d)}_{a,q}$ and
$\C^{(d)}_{b}$. In particular, while the results of
Section~\ref{sec:analysis} have been obtained by assuming a single
coupling $\C_{a,q}^{(d)}$ common to all quarks, using
{\verb NRDD_constraints } such constraints can be generalized to a generic
dependence of such couplings on the flavor $q$.
\section{Conclusions}
\label{sec:conclusions}

Assuming for WIMPs a Maxwellian velocity distribution in the Galaxy we
have explored in a systematic way the relative sensitivity of an
extensive set of existing DM direct detection experiments to each of
the operators $\Q^{(d)}_{a,q}$, $\Q^{(d)}_{b}$ listed in
Eqs.~(\ref{eq:dim5}--\ref{eq:dim7}) up to dimension 7 describing dark
matter effective interactions with quarks and gluons.  In particular
we have focused on a systematic approach, including an extensive set
of experiments and large number of couplings, both exceeding for
completeness similar analyses in the literature.  For all the
operators we have fixed the corresponding dimensional coupling
$\C^{(d)}_{a,q}$ at the scale $\mu_{scale}$=$m_Z$ and used the code
DirectDM~\cite{directdm} to perform the running from $m_Z$ to the
nucleon scale and the hadronization to single-nucleon (N=p,n)
currents, including QCD effects and pion poles that arise in the
nonperturbative matching of the effective field theory to the
low--energy Galilean--invariant nonrelativistic effective theory
describing DM--nucleon interactions. For operators $\Q^{(6)}_{3,q}$
and $\Q^{(6)}_{4,q}$ we have also used the runDM code~\cite{rundm} to
discuss the mixing effect among the vector and axial--vector currents
induced by the running of the couplings above the EW scale, when the
DM vector--axial coupling is assumed to be the same to all quarks.

We find that operators $\Q^{(5)}_{1,q}$, $\Q^{(5)}_{2,q}$,
$\Q^{(6)}_{1,q}$, $\Q^{(6)}_{2,q}$, $\Q^{(7)}_{1}$, $\Q^{(7)}_{2}$,
$\Q^{(7)}_{5,q}$, $\Q^{(7)}_{6,q}$ and $\Q^{(7)}_{10,q}$ take
contributions which correspond to a Spin Independent scaling of the
cross section (possibly combined with explicit dependence from the
exchanged momentum $q^2$ and from the WIMP incoming speed) leading to
an exclusion plot driven by DarkSide--50 at low WIMP mass and XENON1T
at larger $m_{\chi}$. On the other hand for models $\Q^{(6)}_{3,q}$,
$\Q^{(6)}_{4,q}$, $\Q^{(7)}_{3}$, $\Q^{(7)}_{4}$, $\Q^{(7)}_{7,q}$,
$\Q^{(7)}_{8,q}$, and $\Q^{(7)}_{9,q}$ the cross section scaling law
is of the Spin--Dependent type, leading to an exclusion plot driven by
PICASSO and PICO60 (and, sometimes, by CDMSLite) at low WIMP mass, and
by XENON1T at larger WIMP masses.

We also find that for models $\Q^{(7)}_{2}$, $\Q^{(7)}_{3}$,
$\Q^{(7)}_{4}$, $\Q^{(7)}_{6}$, $\Q^{(7)}_{7}$, $\Q^{(7)}_{8}$ and
$\Q^{(7)}_{9}$ the present experimental sensitivity of direct
detection experiments appears not to be able to put bounds consistent
to the validity of the EFT, although only matching the EFT with the
full theory would allow to draw robust conclusions.

The matching between the relativistic effective theory to the NR one
implies a redundancy of the parameters $\C^{(d)}_{a,q}$, implying
that in many cases the DD constraints, that only depend on the ratio
$c_i^n/c_i^p$ between the WIMP--neutron and the WIMP--proton
couplings, can only be discussed for specific benchmarks. In
particular in our exclusion plots we have assumed a
flavor--independent coupling,
$\C^{(d)}_{a,q}$=$\C^{(d)}_{a}$. However, we have shown how, once the
WIMP mass $m_{\chi}$ and the $c_i^n/c_i^p$ ratio are fixed, for all
the $\Q^{(d)}_{a,q}$ models with the exception of $\Q^{(7)}_{7,q}$ and
$\Q^{(7)}_{8,q}$ the expected rate is naturally driven by a dominant
contribution from one of the NR operators ${\cal O}_i$ (possibly
modified by a momentum--dependent Wilson coefficient, ${\cal
  O}_i\rightarrow {\cal O}_iF_i^{\alpha}(q^2)$) without large
cancellations. This implies that the bounds directly obtained within
the context of the NR theory by assuming dominance of one of the
$F_i^{\alpha}(q^2){O}_i$ can be used as discussed in
Ref.~\cite{sogang_scaling_law_nr} to obtain approximate constraints
valid for any choice of the $\Q^{(d)}_{a,q}$ parameters, with an
inaccuracy within a factor of two, but usually smaller. To perform
such task in~\ref{app:program} we provide a simple interpolating
interface in Python. On the other hand, in the case of
$\Q^{(7)}_{7,q}$ and $\Q^{(7)}_{8,q}$ the terms with a
momentum--independent coefficient ${\cal O}_{10}$ and ${\cal O}_{6}$
are next-to-leading order compared to the terms $F_i(q^2){\cal
  O}_{10}$ and $F_i(q^2){\cal O}_{6}$ which depend on the pion and eta
propagators $F_i(q^2)=1/(m_{\pi}^2-q^2)$, $1/(m_{\eta}^2-q^2)$, so
that they cannot be assumed to dominate. Indeed, in this case all the
terms $F_n(q^2){\cal O}_n$ are naturally of the same order with large
cancellations among them.

\appendix

\section{The program}
\label{app:program}
The {\verb NRDD_constraints } code provides a simple interpolating
function written in Python that for a given generalized NR diagonal
term $(F_i^{\alpha})^2{\cal O}_i{\cal O}_i$ among those listed in
Table \ref{table:interferences} (with the exception of those
proportional to a meson pole) calculates the most constraining limit
among the experiments analyzed in~\cite{sogang_scaling_law_nr} on the
effective cross section:

\begin{equation}
\sigma^{\cal N}_{i,\alpha}=\max(\sigma^{p}_{i,\alpha},\sigma^{n}_{i,\alpha}),
  \label{eq:conventional_sigma_nucleon}
\end{equation}

\noindent (for $F_i^{\alpha}$=1,$(q^2)^{-1}$) with:

\begin{equation}
\sigma^{p,n}_{i,\alpha}=(\hat{c}_{i,\alpha}^{p,n})^2\frac{\mu_{\chi{\cal N}}^2}{\pi},
  \label{eq:conventional_sigma}
\end{equation}

as a function of the WIMP mass $m_{\chi}$ and of the ratio
$r_i=\hat{c}_{i,\alpha}^{n}/\hat{c}_{i,\alpha}^{p}$.  The
$\hat{c}_{i,\alpha}^{p,n}$ coefficients are defined in
Eq. (\ref{eq:cr_factorizations}). The code requires the {\verb SciPy }
package and contains only four files, the code
{\verb NRDD_constraints.py }, two data files {\verb NRDD_data1.npy } and
{\verb NRDD_data2.npy }, and a driver template {\verb NRDD_constraints-example.py }.
The module can be downloaded from
\begin{center}
{\verb https://github.com/NRDD-constraints/NRDD } 
\end{center} or cloned by 
\begin{Verbatim}[frame=single,xleftmargin=1cm,xrightmargin=1cm,commandchars=\\\{\}]
  git clone https://github.com/NRDD-constraints/NRDD
\end{Verbatim}
By typing:

\begin{Verbatim}[frame=single,xleftmargin=1cm,xrightmargin=1cm,commandchars=\\\{\}]
  import NRDD_constraints as NR
\end{Verbatim}

\noindent two functions are defined. The function {\verb sigma_nucleon_bound(inter,mchi,r) }
returns the upper bound $(\sigma^{\cal  N}_{i,\alpha})_{lim}$ on the effective cross section of
Eq.(\ref{eq:conventional_sigma_nucleon}) in cm$^2$ as a function of
the WIMP mass {\verb mchi } and of the ratio {\verb r }=$r$ in the
ranges $0.1 \mbox{ GeV}<m_{\chi}<1000$ GeV, $-10^4 <r<10^4$.
The {\verb inter }
parameter is a string that selects the interaction term and
can be chosen in the list provided by the second function
{\verb print_interactions() }:


\begin{Verbatim}[frame=single,xleftmargin=1cm,xrightmargin=1cm,commandchars=\\\{\}]
  NR.print_interactions()\\     
  ['O1_O1','O3_O3', 'O4_O4', 'O5_O5', 'O6_O6',\\ 
   'O7_O7', 'O8_O8', 'O9_O9', 'O10_O10', 'O11_O11',\\
   'O12_O12', 'O13_O13', 'O14_O14', 'O15_O15'\\
   'O5_O5_qm4', 'O6_O6_qm4', 'O11_O11_qm4'] 
\end{Verbatim}

\noindent The list above includes all the $\op_i\op_i F_i^2(q^2)$
terms in Table~\ref{table:interferences} with the exception of those
depending on pion and eta poles which, as explained in
Section~\ref{sec:interferences}, are either subdominant in the case of
models $\Q^{(7)}_3$ and $\Q^{(7)}_4$, or, in the case of $\Q^{(7)}_7$
and $\Q^{(7)}_8$, imply absence of a dominant term altogether (see
Figs.~\ref{fig:epsilon_q7_7} and~\ref{fig:epsilon_q7_8}) so that the
procedure described in this Section leads to an inaccurate estimation
of the constraints. This can be seen explicitly in
Fig.~\ref{fig:q7_78_lambda}, where the red--dashed curve approximates
poorly the bound obtained with a full calculation. The upper bound
returned by {\verb sigma_nucleon_bound(inter,mchi,r) } corresponds to
the results of ~\cite{sogang_scaling_law_nr} with the exception of the
interaction terms with momentum dependence in the Wilson coefficient,
that have been added to include the long--range interactions of
Eq.(\ref{eq:dim5}).

The driver {\verb NRDD_constraints-example.py } calculates the
exclusion plot on the reference cross section:

\begin{equation}
\sigma^{rel}_{ref}=\C^2 \frac{\mu_{\chi{\cal N}}^2}{\pi},
  \label{eq:conventional_sigma_rel}
\end{equation}

\noindent assuming for the interaction between the DM particle $\chi$
and the nucleon $\Psi_N$ the relativistic effective Lagrangian:

\begin{equation}
  {\cal L}=\C \bar{\chi}\gamma^{\mu}\chi \Psi_N\gamma_{\mu}\gamma_5\Psi_N\rightarrow\C\sum_{N=p,n}\left (c^N_7\op_7^N+c^N_9\op_9^N \right ),
  \label{eq:eff_lagrangian}
  \end{equation}

\noindent with $c^p_7$=-2, $c^p_9$=$2 m_N/m_{\chi}$ and
$r_i$=$c^n_i/c^p_i$. Assuming dominance of one operator $\op_i$ at a
time the driver implements a function {\verb coeff(inter,mchi,r) }
that calculates the largest Wilson coefficient between proton and
neutron in absolute value
$max(|c_i^p(m_{\chi},r)|,|c_i^n(m_{\chi},r)|)$ and plots the bound
$(\sigma^{rel}_{ref})_{lim}$ on $\sigma^{rel}_{ref}$
as~\cite{sogang_scaling_law_nr}:
\begin{equation}
(\sigma^{rel}_{ref})_{lim}=\min_{i=7,9} \left(\frac{(\sigma^{\cal
      N}_{i})_{lim}(m_{\chi},r_i)}{max(c_i^p(m_{\chi},r_i)^2,c_i^n(m_{\chi},r_i)^2)}\right ),
\end{equation}

\noindent for the specific choice $r_7$=$r_9$=1. In particular, for a
given value of the WIMP mass {\verb mchi }:

\begin{Verbatim}[frame=single,xleftmargin=1cm,xrightmargin=1cm,commandchars=\\\{\}]
sigma_lim_rel_min=large_number\\
for inter in ['O7_O7','O9_O9']:\\
     c=coeff(inter,mchi,r)\\
     sigma_lim_nucleon_NR=NR.sigma_nucleon_bound(inter,mchi,r)\\
     sigma_lim_rel=sigma_lim_nucleon_NR/c**2\\
     sigma_lim_rel_min=min(sigma_lim_rel_min,sigma_lim_rel)
\end{Verbatim}

\noindent Such limit is also converted into an upper bound
$(\tilde{\Lambda})_{lim}$ in GeV on $\tilde{\Lambda}$ using:

\begin{equation}
(\tilde{\Lambda})_{lim}=\left (\frac{\mu_{\chi{\cal N}} (\hbar
    c)}{\sqrt{(\sigma^{rel}_{ref})_{lim}\pi}}\right)^{\frac{1}{d-4}},
\end{equation}
\noindent with $\hbar c$=1.97$\times 10^{-14}$ cm GeV and $d$=6 the
dimension of the considered operator. An analogous procedure allows to
obtain the red--dashed curve in
Figs.~\ref{fig:q5_lambda}--\ref{fig:q6_34_lambda} and
Figs.~\ref{fig:q7_12_lambda}--\ref{fig:q7_9_10_lambda} that reproduce
accurately most of the results of Section~\ref{sec:analysis}.

\section*{Acknowledgements}
This research was supported by the Basic Science
Research Program through the National Research Foundation of
Korea~(NRF) funded by the Ministry of Education, grant number
2016R1D1A1A09917964.









\end{document}